\documentclass[letterpaper,twocolumn,10pt]{article}

\usepackage{usenix-2020-09}

\usepackage{url} 
\usepackage{framed}
\usepackage{mdframed}
\usepackage{listings}
\usepackage{multicol}
\usepackage{amssymb}
\usepackage{lipsum}
\usepackage{adjustbox}
\usepackage[font=bf,skip=\baselineskip]{caption}
\usepackage{tabularx}
\usepackage{seqsplit}
\usepackage{multirow}
\usepackage{booktabs}
\usepackage{graphicx}
\usepackage{hyperref}
\usepackage{subcaption}
\usepackage{makecell}
\usepackage{xcolor}
\usepackage{colortbl}
\usepackage[ruled,vlined,linesnumbered] {algorithm2e}
\usepackage{amsmath}
\usepackage{tikz}
\usepackage{tcolorbox}
\usepackage{threeparttable}
\usepackage{placeins}
\usepackage{makecell}
\usepackage{longtable}
\usepackage{float}

\usepackage{boldline}
\usepackage[ruled,vlined]{algorithm2e}
 
\usepackage{pifont}% 
\usepackage{balance}
\let\labelindent\relax
\usepackage{enumitem}
\usepackage[htt]{hyphenat}
\usepackage{lineno}
\usepackage{tabularx}
\usepackage{booktabs}
\usepackage{amssymb}

\definecolor{customblue}{HTML}{006ca6}
\definecolor{customgreen}{HTML}{009264}
\definecolor{custombrown}{HTML}{ff3d00}
\AtEndPreamble{

 \hypersetup{
  colorlinks = true,
  linkcolor = customgreen,
  anchorcolor = purple,
  citecolor = customgreen,
  filecolor = purple,
  urlcolor = customblue
 }
}

\newcommand{\bench}{\texttt{VoiceWukong}}

\begin{document}

\title{\bench{}: Benchmarking Deepfake Voice Detection}

\author{
{\rm Ziwei Yan}\footnotemark[1]\\
Huazhong University of Science and Technology \\
yandoit@hust.edu.cn
\and
{\rm Yanjie Zhao}\thanks{Ziwei Yan and Yanjie Zhao contributed equally to this work.}\\
Huazhong University of Science and Technology \\
yanjie\_zhao@hust.edu.cn
\and
{\rm Haoyu Wang}\thanks{Haoyu Wang is the corresponding author (haoyuwang@hust.edu.cn).}\\
Huazhong University of Science and Technology \\
haoyuwang@hust.edu.cn
}

\maketitle

\begin{abstract}

With the rapid advancement of technologies like text-to-speech (TTS) and voice conversion (VC), detecting deepfake voices has become increasingly crucial. However, both academia and industry lack a comprehensive and intuitive benchmark for evaluating detectors. Existing datasets are limited in language diversity and lack many manipulations encountered in real-world production environments.

To fill this gap, we propose \bench{}, a benchmark designed to evaluate the performance of deepfake voice detectors. To build the dataset, we first collected deepfake voices generated by 19 advanced and widely recognized commercial tools and 15 open-source tools. We then created 38 data variants covering six types of manipulations, constructing the evaluation dataset for deepfake voice detection. \bench{} thus includes 265,200 English and 148,200 Chinese deepfake voice samples. Using \bench{}, we evaluated 12 state-of-the-art detectors. \texttt{AASIST2} achieved the best equal error rate (EER) of 13.50\%, while all others exceeded 20\%. Our findings reveal that these detectors face significant challenges in real-world applications, with dramatically declining performance. In addition, we conducted a user study with more than 300 participants. The results are compared with the performance of the 12 detectors and a multimodel large language model (MLLM), i.e., \texttt{Qwen2-Audio}, where different detectors and humans exhibit varying identification capabilities for deepfake voices at different deception levels, while the LALM demonstrates no detection ability at all.
Furthermore, we provide a \textbf{leaderboard} for deepfake voice detection, publicly available at \url{https://voicewukong.github.io}.

\end{abstract}

\section{Introduction}
\label{sec:introduction}
The rapid development of technologies such as TTS (Text-to-Speech) and VC (Voice Conversion) has brought great convenience to people in areas like entertainment and accessibility services. However, it is a double-edged sword. Illegal actors may exploit deepfake voices for various criminal activities. For example, in 2019, criminals used AI-based software to impersonate the voice of a U.K.-based energy firm's chief executive and requested a fraudulent transfer of €220,000~\cite{noauthor_2019-mu}. 
To counter the growing threats posed by deepfake voice technology, researchers are actively developing deepfake voice detection methods and creating open-source datasets for evaluation. 
For instance, traditional pipeline detection methods like \texttt{LFCC-LCNN}~\cite{wang2021comparative}, the one-class learning-based detection method~\cite{Zhang_2021}, emerging end-to-end detection models such as \texttt{AASIST}~\cite{aasist} and \texttt{RawNet2}~\cite{tak2021end}, as well as open-source deepfake voice datasets like ASVspoof~\cite{2019asvspoof,2021asvspoof}, have all garnered widespread attention.

Unfortunately, academic deepfake voice detection methods often excel on specific datasets but fall short in real-world scenarios~\cite{in-the-wild}. The rise of commercial tools and the latest generative models has produced increasingly convincing synthetic voices, outpacing current detection capabilities~\cite{patel2023detectorbehindgeneration}. A key issue is the reliance on outdated or generic datasets for evaluation, which fail to reflect the sophistication of modern deepfake technologies.
In practical applications, detectors struggle with poor generalization to unknown attacks and lack large-scale in-the-wild datasets~\cite{23survey-yi}. Additionally, most methods focus solely on original content, overlooking the impact of post-processing manipulations (such as noise injection) on detection accuracy~\cite{wu2024clad}.
Given these challenges, there is a pressing need for a comprehensive benchmark to objectively evaluate various detection methods, thereby \textbf{bridging the gap between academic research and real-world applications}.

To address this gap, we introduce \bench{}, a comprehensive deepfake voice detection benchmark that incorporates various voice manipulations. \bench{} focuses on English and Chinese, the two most widely spoken languages globally, and features voices synthesized by advanced commercial tools and open-source models.
We evaluated 12 state-of-the-art detection models, visually presenting their performance differences. Our fine-grained analysis of detector performance across different manipulations reveals potential avenues for optimization and improvement.
Recognizing that \textbf{humans are the primary targets of deepfakes}, we conducted a user study involving over 300 participants. Based on the results, we classified deepfake voices into three levels. We then analyzed the performance of the detectors at each level, comparing the detection capabilities of users versus automated systems for these synthesized voices.

Our main contributions can be summarized as follows:
\begin{itemize}

\item  \textbf{A dataset addressing gaps.} Our dataset encompasses both English and Chinese languages, leveraging 19 advanced commercial tools and 15 open-source models. Through six types of manipulations, it has accumulated 38 data variants, resulting in a total of 265,200 English and 148,200 Chinese deepfake voice samples. To our knowledge, \textbf{this dataset is the first to extensively incorporate manipulation variants and compile the largest collection of commercially generated voice samples}.

\item \textbf{A comprehensive benchmark.} 
We evaluated 12 advanced deepfake voice detectors using \bench{}. Results show that most detectors have an equal
error rate (EER) above 20\%, with three detectors exhibiting random performance on either the Chinese or English dataset. \texttt{AASIST2}~\cite{tak2022aasist2} achieved the best EER (13.50\% for English and 13.54\% for Chinese), yet this falls significantly short of the 0.82\% EER reported on its original evaluation dataset, underscoring the challenges these detectors face in real-world applications. We further compared detector performance and conducted a fine-grained analysis to identify specific manipulations that cause performance degradation for each detector.

\item \textbf{A large-scale user study.} We conducted a user study involving over 300 participants to categorize deepfake voices into three levels of increasing difficulty (levels 0-2) based on their actual deception effectiveness. We then evaluated the performance of the 12 detectors and a multimodel large language model (MLLM), \texttt{Qwen2-audio}~\cite{chu2024qwen2}, across these levels. Results show that humans have false acceptance rates (FARs) of 18.97\% for level 0 deepfakes in English and 4.20\% in Chinese, outperforming most detectors. For level 2 deepfake voices, human FARs exceed 82\% in both languages, falling behind most detectors. \texttt{Qwen2-audio} has an F1-Score of zero on the English dataset, indicating its inability to detect deepfake voices. We also examined human-focused features in deepfake voice detection to enhance detector-human collaboration in identifying synthetic audio.

\end{itemize}

\section{Background and Related Work}
\label{sec:background}

\begin{table*}[ht!]
\caption{The details of commonly used deepfake voice datasets. Note that, various datasets have issues such as single language and lack of manipulations.}
\label{tab:dataset_detail}
\centering
\footnotesize
\setlength{\tabcolsep}{2pt}
\begin{tabularx}{\textwidth}{@{}X|X|XX|X|XXX|XXX|X@{}}
\toprule
\multicolumn{1}{c|}{\textbf{Dataset}}& \multicolumn{1}{c|}{\textbf{FoR}} & \multicolumn{2}{c|}{\textbf{ASVspoof 2019}} & \multicolumn{1}{c|}{\textbf{WaveFake}}& \multicolumn{3}{c|}{\textbf{ASVspoof2021}} & \multicolumn{3}{c|}{\textbf{ADD 2022}} & \multicolumn{1}{c}{\textbf{In-the-Wild}}   \\
\midrule
\multicolumn{1}{c|}{Year} & \multicolumn{1}{c|}{2019} & \multicolumn{2}{c|}{2019} & \multicolumn{1}{c|}{2021} & \multicolumn{3}{c|}{2021} & \multicolumn{3}{c|}{2022} & \multicolumn{1}{c}{2022} \\
\multicolumn{1}{c|}{Language} & \multicolumn{1}{c|}{English} & \multicolumn{2}{c|}{English} & \multicolumn{1}{c|}{English \& Japanese} & \multicolumn{3}{c|}{English} & \multicolumn{3}{c|}{Chinese} & \multicolumn{1}{c}{English} \\
\multicolumn{1}{c|}{Corpus} & \multicolumn{1}{c|}{A phrase dataset~\cite{fratxt}} & \multicolumn{2}{c|}{VCTK~\cite{vtck}} & \multicolumn{1}{c|}{LJSpeech~\cite{ljspeech} \& JSUT~\cite{sonobe2017jsut}} & \multicolumn{3}{c|}{VCTK \& Other\textsuperscript{1}} & \multicolumn{3}{c|}{AI-1,3,4~\cite{bu2017aishell,shi2020aishell,fu2021aishell}\textsuperscript{2}} & \multicolumn{1}{c}{-} \\
\midrule
\multicolumn{1}{c|}{Subset} & \multicolumn{1}{c|}{-} & \multicolumn{1}{c}{LA} & \multicolumn{1}{c|}{PA} & \multicolumn{1}{c|}{-} & \multicolumn{1}{c}{LA} & \multicolumn{1}{c}{PA} & \multicolumn{1}{c|}{DF} & \multicolumn{1}{c}{LF} & \multicolumn{1}{c}{PF} & \multicolumn{1}{c|}{FG-D} & \multicolumn{1}{c}{-}  \\
\multicolumn{1}{c|}{Types} & \multicolumn{1}{c|}{TTS} & \multicolumn{1}{c}{TTS,VC} & \multicolumn{1}{c|}{Replay} & \multicolumn{1}{c|}{TTS} & \multicolumn{1}{c}{TTS,VC} & \multicolumn{1}{c}{Replay} & \multicolumn{1}{c|}{TTS,VC} & \multicolumn{1}{c}{TTS,VC} & \multicolumn{1}{c}{PF\textsuperscript{3}} & \multicolumn{1}{c|}{TTS,VC} & \multicolumn{1}{c}{TTS}  \\
\multicolumn{1}{c|}{Goal} & \multicolumn{1}{c|}{DD} & \multicolumn{1}{c}{ASV} & \multicolumn{1}{c|}{ASV} & \multicolumn{1}{c|}{DD} & \multicolumn{1}{c}{ASV} & \multicolumn{1}{c}{ASV} & \multicolumn{1}{c|}{DD} & \multicolumn{1}{c}{DD} & \multicolumn{1}{c}{DD} & \multicolumn{1}{c|}{DD} & \multicolumn{1}{c}{DD} \\
\multicolumn{1}{c|}{Manipulation} & \multicolumn{1}{c|}{-} & \multicolumn{1}{c}{-} & \multicolumn{1}{c|}{-} & \multicolumn{1}{c|}{-} & \multicolumn{1}{c}{Trans\textsuperscript{4}} & \multicolumn{1}{c}{Noisy, Reverb} & \multicolumn{1}{c|}{-} & \multicolumn{1}{c}{Noisy} & \multicolumn{1}{c}{-} & \multicolumn{1}{c|}{-} & \multicolumn{1}{c}{Noisy} \\
\multicolumn{1}{c|}{Commercial} & \multicolumn{1}{c|}{6} & \multicolumn{1}{c}{0} &\multicolumn{1}{c|}{-} & \multicolumn{1}{c|}{0} & \multicolumn{1}{c}{0} & \multicolumn{1}{c}{-} & \multicolumn{1}{c|}{0} & \multicolumn{1}{c}{-} & \multicolumn{1}{c}{-} & \multicolumn{1}{c|}{-} & \multicolumn{1}{c}{-} \\
\multicolumn{1}{c|}{Academic} & \multicolumn{1}{c|}{1} & \multicolumn{1}{c}{17} & \multicolumn{1}{c|}{-} & \multicolumn{1}{c|}{7} & \multicolumn{1}{c}{13} & \multicolumn{1}{c}{-} & \multicolumn{1}{c|}{>100} & \multicolumn{1}{c}{-} & \multicolumn{1}{c}{-} & \multicolumn{1}{c|}{-} & \multicolumn{1}{c}{-} \\
\bottomrule
\end{tabularx}
\parbox{\textwidth}{\footnotesize 1: Other undisclosed corpora.}
\parbox{\textwidth}{\footnotesize 2: AI-1, AI-3, and AI-4 represent AISHELL-1, AISHELL-3, and AISHELL-4, respectively.}
\parbox{\textwidth}{\footnotesize 3: PF represents partially fake that generated by manipulating only a few words in the original bonafide utterances with real or synthesized voices.}
\parbox{\textwidth}{\footnotesize 4: The LA part of ASVspoof2021 transmitted the original voice through various telephone systems.}
\end{table*}

\subsection{Deepfake Voice Detection}

Deepfake voice detectors primarily fall into two categories: traditional pipeline detectors and the increasingly researched end-to-end detectors~\cite{23survey-yi}. The pipeline consists of a frontend feature extractor and a backend classifier. The classifier determines authenticity based on the features extracted by the feature extractor. Common features include spectral features represented by mel frequency cepstral coefficient (MFCC)~\cite{chen2010speaker}, linear frequency cepstral coefficients (LFCC)~\cite{tian2016spoofing}, constant-Q transform (CQT)~\cite{cheng2019replay}; supervised embedding features~\cite{pan2022speaker}; and self-supervised embedding features represented by Wav2vec based features~\cite{tak2022aasist2}, XLS-R~\cite{babu2021xls} based features~\cite{martin2022vicomtech}. Moreover, researchers have also attempted to explore some non-traditional features. Wang et al.~\cite{wang2019voicepop} analyzed pop noise from close microphone speaking to detect deepfake voices. Doan et al.~\cite{doan2023bts} detected deepfakes by evaluating the correlation between breathing, talking (speaking), and silence sounds. Common backend classifiers include traditional classifiers represented by GMM-based classifiers~\cite{de2012synthetic}, and deep learning classifiers represented by ResNet~\cite{he2016deep} based classifiers~\cite{tomilov2021stc}, Res2Net~\cite{gao2019res2net} based classifiers~\cite{kim2023phase}, and DARTS~\cite{liu2018darts} based classifiers~\cite{ge2021partially}.

End-to-end detectors have also garnered widespread attention in the research community. \texttt{RawNet2}~\cite{jung2020improved} is a neural network designed for end-to-end speech recognition and speaker verification that directly processes raw audio waveforms. Tak et al.~\cite{tak2021end} were the first to apply \texttt{RawNet2} to anti-spoofing. Wang et al.~\cite{wang2022audio} proposed a joint optimization approach based on the weighted additive angular margin loss to extend and optimize the \texttt{RawNet2} based deepfake voice detector. Tak et al.~\cite{rawgatst} leveraged the merit of graph attention networks (GATs) to learn the relationships between cues located in different sub-bands or different temporal intervals~\cite{tak2021graphattentionnetworksantispoofing}, proposing \texttt{RawGAT-ST}, which achieved excellent performance on ASVspoof2019. \texttt{RawGAT-ST} uses a pair of parallel graphs to simultaneously model temporal and spectral information, then combines them with element-wise multiplication. Jung et al.~\cite{aasist} proposed integrating these two heterogeneous graphs with heterogeneity-aware techniques and developed the \texttt{AASIST}, which achieved superior performance on ASVspoof2019.

Unfortunately, most existing detectors are limited to pursuing performance on a single dataset, neglecting many challenges encountered in real-world applications. Ba et al.~\cite{language-across} highlighted these limitations in cross-language detection and proposed adaptation strategies. Zhang et al.~\cite{keeplearning} pointed out the insufficiency of models in adapting to unknown new attacks and conducted research on continual learning in deepfake voice detection. Wang et al.~\cite{deepsonar} and Wu et al.~\cite{wu2024clad} considered the impact of manipulations on detector robustness, an aspect that most people have not taken into account.

\subsection{Benchmarks and Datasets}

\noindent \textbf{Benchmarks.} 
Recent research in deepfake detection has primarily focused on benchmarking face detection methods. Notable contributions include the CDDB benchmark by Li et al.\cite{Li_2023_WACV}, which simulates real-world scenarios, and the comprehensive evaluation by Deng et al.\cite{towardsbenchmark} using multiple generation methods and detection metrics. Pei et al.~\cite{pei2024deepfakegenerationdetectionbenchmark} provided a thorough survey and evaluation of deepfake face generation and detection techniques across various datasets and sub-fields.
In the voice domain, Zang et al.~\cite{zang2024ctrsvddbenchmarkdatasetbaseline} introduced CtrSVDD, a large-scale benchmark for detecting singing voice synthesis models. \textbf{To the best of our knowledge, \bench{} is the first comprehensive and in-depth benchmark focusing on deepfake voice detection.}

\noindent \textbf{Datasets.} The commonly used evaluation datasets for deepfake voice detectors include FoR~\cite{FoR}, ASVspoof2019~\cite{2019asvspoof},  WaveFake~\cite{frank2021wavefake}, ASVspoof2021~\cite{2021asvspoof}, ADD2022~\cite{2022add}, and In-the-Wild~\cite{in-the-wild}, as shown in \autoref{tab:dataset_detail}. 
ASVspoof2019 is constructed for \textit{automatic speaker verification (ASV)} tasks and includes a replay subset (ASVspoof2019-PA). It has received a lot of attention in the field of \textit{deepfake detection (DD). }ASVspoof2021 builds upon ASVspoof2019 by adding a section specifically for deepfake detection (ASVspoof2021-DF). Only WaveFake is constructed across multiple languages, but it does not focus on the most widely used languages. In-the-wild has a limited scope, focusing only on deepfake voices of celebrities and politicians. FoR is derived from seven open-source and commercial methods. Only a few datasets include manipulations like noise (ASVspoof2021, ADD2022, In-the-wild) and replay attacks (ASVspoof2019, ASVspoof2021). Overall, our dataset offers the broadest coverage of commercial tools, encompasses the widest range of voice manipulation variants, and targets the most representative languages.

\subsection{Threat Model}

The threat model in this study focuses on the malicious use of deepfake voice technology such as fraud and impersonation. Adversaries are assumed to have access to advanced commercial and open-source voice synthesis tools, enabling them to generate highly convincing synthetic voices in English and Chinese, and employ post-processing techniques to enhance realism and evade detection.
The rapid advancement of voice synthesis technology presents a growing threat, often outpacing the development of detectors~\cite{patel2023detectorbehindgeneration}. Current academic detectors may fail in the real world due to poor generalization and outdated datasets, creating a gap between lab results and practical effectiveness against sophisticated deepfake voices. Our goal is to bridge this gap by providing a comprehensive benchmark that reflects real-world threats, evaluates state-of-the-art detection methods, and incorporates human perception in assessing deepfake voice detection effectiveness.

%  The threat is compounded by the rapid advancement of voice synthesis technology, which often outpaces the development of detection methods~\cite{~\cite{patel2023detectorbehindgeneration}}. Additionally, the model considers that current academic detection approaches may struggle with real-world applications due to poor generalization and reliance on outdated datasets, creating a significant gap between laboratory performance and practical effectiveness against sophisticated deepfake voices. Our goal is to
% bridge this gap by providing a comprehensive benchmark that
% reflects real-world threats, evaluates state-of-the-art detection
% methods, and incorporates human perception in assessing
% deepfake voice detection effectiveness.

\begin{figure*}[ht!]
    \centering
    \includegraphics[width=\textwidth]{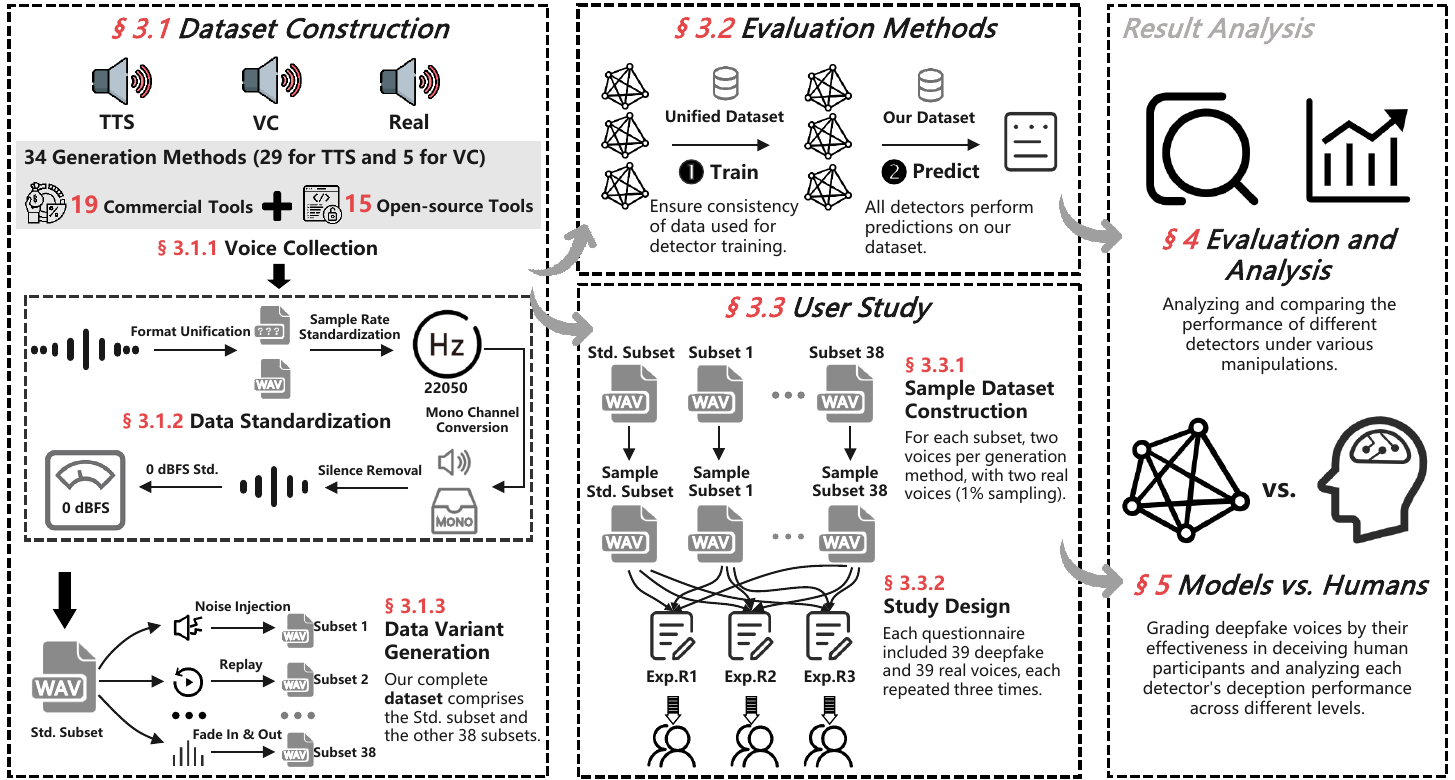}
    \caption{The overall workflow of the \bench{} benchmark construction.}
    \label{fig:overview}
\end{figure*}

\section{Benchmark Construction}
\label{sec:overview}

In this section, we introduce the construction process of \bench{}, as illustrated in \autoref{fig:overview}. \autoref{sec:datasetconstruction} introduces the dataset construction process. \autoref{sec:evamethods} presents our unified training and evaluation of detectors. \autoref{sec:userstudy} details our large-scale user study. Finally, \autoref{sec:metrics} discusses the evaluation metrics used in \bench{}.

\subsection{Dataset Construction}
\label{sec:datasetconstruction}
\subsubsection{Voice Collection}
\label{sec:voices collection}
\noindent \textbf{Generation Methods.} Since TTS and VC are mainstream methods for generating deepfake voices~\cite{Audiosurvey}, our dataset focuses on these two types. We collected 15 open-source generation models that are either prominent in the research field or have the highest star ratings on GitHub. Given that adversaries might use commercial tools to synthesize deepfake voices in real-world scenarios, we additionally collected 19 such tools capable of generating deepfake voices for research purposes and paid the necessary fees for their use. 
\textbf{We carefully examined the terms of service for each commercial tool to confirm their permissibility for research purposes.} \bench{} is a non-commercial resource, thereby safeguarding against any potential infringement of intellectual property rights.
To our knowledge, our dataset involves the most extensive range of commercial tools. 
The 34 methods (29 for TTS and 5 for VC) are listed in \autoref{tab:genmethod} in the Appendix, all supporting English and 19 supporting Chinese.

\noindent \textbf{Generation Process.} 
\bench{} encompasses English and Chinese, the two most widely used languages globally. Deepfake voice generation utilizes the English VCTK~\cite{vtck} and Chinese MAGICDATA~\cite{magicdata2019} datasets. VCTK is widely used in voice cloning and conversion research, featuring diverse English textual content. MAGICDATA is a large-scale Mandarin speech dataset with extensive read speech data, covering domains like news, dialogues, and question-answering. After deduplication and length filtering (5-30 words for English, 5-35 characters for Chinese), we selected 100 sentences each from VTCK and MAGICDATA for fixed-text generation, and extracted 3,400 English and 1,900 Chinese sentences for random-text generation. For each method, we produced 200 deepfake voices per supported language: 100 with fixed text and 100 with random text. Methods with 10 or more speakers yielded 10 fixed-text and 10 random-text voices per speaker from 10 speakers. For methods with fewer speakers, we distributed the 200 voices evenly among available speakers. Generation was automated for open-source models and API-based tools, while web-based tools required manual input. For the five open-source VC models, we provided corresponding original voices. In total, we collected 6,800 English deepfake voices across 34 methods and 3,800 Chinese deepfake voices across 19 Chinese-supporting methods.

\noindent \textbf{Real Voice Collection.} We also collected real voices from the VCTK and MAGICDATA datasets. To ensure data balance, we randomly selected text content not present in any deepfake voices and collected an equal number of real voices: 6,800 from VCTK and 3,800 from MAGICDATA.

\subsubsection{Data Standardization}
\label{sec:datastd}

Inspired by previous research~\cite{FoR}, the diversity of our dataset sources necessitates data standardization to eliminate biases.
We implemented a standardized preprocessing pipeline for all voice samples, which included converting files to WAV format, resampling to 22,050 Hz, transforming to Mono channel, trimming silent segments from the beginning and end, and normalizing the volume to 0 dBFS.

\noindent \textbf{Format Unification.} Our collection methods include various open-source models and commercial tools, resulting in non-uniform audio formats, mainly WAV and MP3. To eliminate format bias, we used \texttt{pydub}\footnote{\url{https://github.com/jiaaro/pydub}} to convert all audio files to WAV format, as it supports conversion between multiple formats with minimal loss.

\noindent \textbf{Sample Rate Standardization.} Regarding sample rates, the voices from different sources also vary. To evaluate the impact of sample rate changes on detectors, we standardized all data to a sample rate of 22,050~Hz. This rate captures frequencies up to 11,025~Hz, which is sufficient for the main spectral range of human speech (300~Hz to 3,400~Hz). It provides adequate frequency resolution while saving space and improving processing efficiency compared to higher rates like 44,100~Hz and 48,000~Hz. We used the commonly applied  \texttt{librosa}\footnote{\url{https://github.com/librosa/librosa}} for resampling all voice files to 22,050~Hz.

\noindent \textbf{Mono Channel Conversion.} Since the VCTK dataset uses two microphones for recording, we consistently used its data from \texttt{Microphone~2}. Additionally, the channel settings vary across different audio sources. To avoid the impact of different channels on the detectors, we used \texttt{librosa} to convert all voice files to Mono, eliminating bias.

\noindent \textbf{Silence Removal.} Real speech recordings often include silent segments at the beginning and end, which speech synthesis methods may not always replicate. To eliminate bias, we removed these silent parts from all voice files. A Python script was used to calculate the smoothed energy envelope of each voice, removing segments below the 20th percentile at the beginning and end.

\noindent \textbf{Volume Normalization to 0 dBFS.} When collecting voices from various commercial tools, some offer volume adjustment features while others do not. Moreover, the volume levels of voices generated by open-source models may vary. To eliminate the impact of volume differences on detection results, we normalize all voices to 0 dBFS.

\subsubsection{Data Variant Generation}
To cover various real-world scenarios that detectors might encounter, we applied additional manipulations to the standardized datasets. These manipulations reflect basic techniques malicious actors could use on deepfake voices and include six types: noise injection (NI), volume control (VC), time stretching (TS), sample rate changes (SR), replay (RE), and fade in \& out (FD) effects. Except for replay, each type underwent fine-grained manipulations to create multiple variants.

\noindent \textbf{Noise Injection~\cite{FoR,noiseij1}.} ESC-50~\cite{piczak2015esc} is a widely recognized environmental sound classification dataset that divides sounds into five categories: animal sounds, natural soundscapes and water sounds, human (non-speech) sounds, interior/domestic sounds, and exterior/urban noises, each containing 400 five-second recordings. For each voice in our standardized dataset, we randomly selected a noise audio from each category and injected noise at Signal-to-Noise Ratios (SNRs) of 15 dB, 20 dB, and 25 dB. We also added Gaussian white noise to each voice using the same SNR settings.

\noindent \textbf{Volume Control~\cite{wilmering2020history}.} To examine if varying volume levels affect deepfake voice detection, we applied multiple volume adjustments to the standardized dataset. Since we were unaware of the adjustment outcomes, we conducted listening tests to ensure the voice content remained clearly audible. We set the lowest volume at 0.5 times and the highest at 1.5 times the standardized dataset, with intermediate levels at 0.75 and 1.25 times the original volume.

\noindent \textbf{Time Stretching~\cite{ts}.} Time stretching is a technique that changes the playback speed of audio without altering its other attributes, effectively adjusting the speech rate. We applied time stretching to the standardized dataset at 0.8, 0.9, 1.1, and 1.2 times the original speed to examine the impact of accelerated and decelerated playback on deepfake voice detection.

\noindent \textbf{Voice Resampling.} Generally, a higher sample rate~\cite{samplerate} provides greater detail and quality in voice, making it sound more realistic and natural. However, higher sample rates also increase storage requirements. Therefore, generation tools often balance sample rates based on their needs. Lower sample rates lose some high-frequency information, potentially impacting detectors that rely on this data. We previously set 22,050~Hz as the sample rate for our standardized dataset and further resampled it at higher rates of 32,000~Hz and 44,100~Hz.

\noindent \textbf{Replay~\cite{wilmering2020history}.} Replay 
% ,janicki2016assessment
attacks~\cite{villalba2010speaker} have long been exploited as a simple yet highly challenging method to test system robustness. To assess the robustness of different detectors against replay attacks, we re-recorded the standardized dataset. We played all voices at maximum volume on a Lenovo Xiaoxin Pro14 2023 laptop and recorded them simultaneously with a Deli 14870 omnidirectional microphone placed 1 meter away at the same height. The recordings were conducted in a quiet, unoccupied indoor environment.

\noindent \textbf{Fade In \& Out.} Fade in and fade out are common voice editing techniques that create smooth transitions and more natural voice connections~\cite{fadein,fadeout}. Fade in gradually increases the volume from zero at the beginning of a voice clip, while fade out gradually decreases it to zero at the end, creating a smooth transition and more natural voice connections. This is a common voice editing technique. We applied fade in \& out to each voice in the standardized dataset using three fade shapes: linear, logarithmic, and exponential at ratios of 0.1, 0.2, and 0.3.

\begin{table}[]
\centering
\caption{Overview of our dataset. Within the deepfake voices, random and fixed text accounts for half.}
\label{tab:datasetoverview}
\resizebox{\columnwidth}{!}{%
\begin{tabular}{|c|c|c|c|c|c|}
\hline
\multirow{3}{*}{\textbf{English Voices}} & \multirow{3}{*}{530,400} & \multirow{2}{*}{\textbf{Fake Voices}} & \multirow{2}{*}{265,200} & \textbf{Fixed-text Voices} & 132,600 \\
\cline{5-6}
 &  &  &  & \textbf{Random-text Voices} & 132,600 \\
\cline{3-6}
 &  & \textbf{Real Voices} & 265,200 & \multicolumn{2}{c|}{-} \\
\hline
\multirow{3}{*}{\textbf{Chinese Voices}} & \multirow{3}{*}{296,400} & \multirow{2}{*}{\textbf{Fake Voices}} & \multirow{2}{*}{148,200} & \textbf{Fixed-text Voices} & 74,100 \\
\cline{5-6}
 &  &  &  & \textbf{Random-text Voices} & 74,100 \\
\cline{3-6}
 &  & \textbf{Real Voices} & 148,200 & \multicolumn{2}{c|}{-} \\
\hline
\multicolumn{1}{|c|}{\textbf{Total}} & \multicolumn{1}{c|}{826,800} & \multicolumn{4}{c|}{-}\\
\hline
\end{tabular}%
}
\end{table}

Using six types of manipulations, \textbf{we generated 38 variants of the standardized (std.) subset, resulting in a total of 39 subsets}. As shown in \autoref{tab:datasetoverview}, our dataset construction process yielded 265,200 English and 148,200 Chinese deepfake voices, along with an equal number of real voices.

\subsection{Evaluation Methods}
\label{sec:evamethods}

We now describe the evaluation of existing deepfake voice detectors using our constructed dataset. Our detector selection process is detailed in \autoref{sec:evaluatedmethods}. All detectors are trained on the ASVspoof2019-LA dataset (if necessary). We use published pre-trained models when available, or retrain models using the original authors' hyperparameters and settings, selecting the best-performing checkpoint for evaluation. This approach preserves each detector's optimal performance as determined by its creators.
During evaluation, we maintain the input sequence length specified by the original authors, rather than fixing the voice sampling duration, due to differences in sample rates between our dataset and ASVspoof2019. This ensures consistency between training and evaluation sequence lengths, enabling fair comparisons across detectors.
This methodology allows us to compare each detector's performance while maintaining its peak capabilities and avoiding inconsistencies that could arise from fixed sampling durations.

\subsection{User Study}
\label{sec:userstudy}
To examine the real-world deception effectiveness of deepfake voices generated by the selected methods and manipulation variants, we conducted a user study involving 318 participants. Based on the deception performance of these voices in the study, we divided the dataset into three levels to enable further evaluation of detectors in \autoref{sec:modelvshumanresult}. This section outlines the data sampling, study design, and grading method, while \textbf{\underline{ethical considerations and participant details} are provided in Appendix \autoref{sec:ethics} and \autoref{sec:Participants}}.

\subsubsection{Sample Dataset Construction} 
\label{sec:study data sampling}

Given our dataset's large scale, we sampled 1\% for the user study, balancing practicality with comprehensive representation. We focused on the random text portion to avoid potential bias from repeated content in the fixed text samples.  As illustrated in \autoref{fig:overview}, our sampling process was: 1) For each subset, we randomly selected two random-text deepfake voices per generation method, ensuring unique texts across subsets. 2) We then sampled an equal number of real voices from the same subset. 
This approach resulted in 38 Chinese deepfake and 38 real voices (corresponding to the 19 generation methods supporting Chinese), as well as 68 English deepfake and 68 real voices (34 generation methods) per subset. With a total of 39 subsets, our final \textbf{sample dataset} comprised 2,964 Chinese voices and 5,304 English voices.

\subsubsection{Study Design}
\label{sec:studydesign}

\noindent \textbf{Questionnaire Setup.} 
To maintain full control over the dataset, we set up a questionnaire on our server and provided each participant with a unique link to access the survey. At the start, users were required to read instructions that explicitly directed them to use headphones to minimize environmental noise and interference. Participants could only proceed after reading, understanding, and agreeing to these instructions. Furthermore, the response time for the entire questionnaire was strictly limited. Questionnaires completed in less than 25 minutes or more than two hours were considered invalid.
For each subset within the \textbf{sample dataset}, we randomly paired one deepfake voice with one real voice.
To construct each questionnaire, we randomly selected one unused pair from each subset, resulting in 78 voices per questionnaire. This ensured that every voice from the sample dataset appeared in each questionnaire round.
After one round of questionnaire generation, we produced 38 different questionnaires for the Chinese dataset and 68 for the English dataset.
We conducted three rounds of data collection, ultimately generating 114 Chinese questionnaires and 204 English questionnaires.

\noindent \textbf{Tasks.}     
The user study is structured around three distinct tasks. In the first task, participants are presented with randomly selected voice samples and asked to determine their origin. As illustrated in \autoref{fig:questionnaire} in the Appendix, users listen to each voice and must categorize it as either ``Human'' or ``Not human''. They are required to make a selection before proceeding and are not permitted to revise their previous choices.
Upon completion of the initial assessment, the second task commences. Participants revisit each deepfake voice sample, with their earlier judgments displayed as a reminder. They are then tasked with evaluating the generation quality on a three-point scale: \texttt{1} indicates an easily detectable deepfake, \texttt{2} suggests a deepfake identifiable upon close scrutiny, and \texttt{3} represents a voice indistinguishable from a genuine human recording. Following this evaluation, participants are prompted to identify the factors that influenced their decision-making process. They can select from predefined options such as volume and background noise or provide custom responses.
The third task serves as an attention check for participants. Additionally, attention tests are embedded once within both the first and second tasks. These tests require participants to listen to a specific recording and answer content-related questions. Incorrect responses result in immediate termination of the questionnaire to prevent random guessing, and these participants are subsequently disqualified from the study.

\noindent \textbf{Deepfake Voice Deception Grading.} 
We developed a systematic approach to grade deepfake voices based on their effectiveness in deceiving human participants. The grading process for deepfake voices produced by a specific tool under particular manipulation conditions was as follows.
Our study consisted of three experimental rounds (as illustrated in \autoref{fig:overview}), each considered a separate experiment within our overall investigation. For each round, we collected responses from 38 Chinese and 68 English questionnaires.
As noted in \autoref{sec:study data sampling}, each generation method was represented by two deepfake voice samples from each subset.
We analyzed the deception outcomes for each experimental round, classifying them into three distinct scenarios \textbf{corresponding to different deception levels (0 to 2)}:  no voices successfully deceiving humans (level 0), one voice successfully deceiving humans (level 1), or two voices successfully deceiving humans (level 2).
The final grade for a generation method under specific conditions was determined by the majority outcome across the three experimental rounds. In cases of evenly distributed results (i.e., one round each at levels 0, 1, and 2, regardless of order), we assigned a final level of 1.

\subsection{Metrics}
\label{sec:metrics} 

In \bench{}, we employ a comprehensive set of quantitative metrics to analyze the performance of various detectors. These metrics include False Acceptance Rate (FAR), False Rejection Rate (FRR), Equal Error Rate (EER)~\cite{eerfarfrr}, Accuracy (ACC), F1-score, and Area Under the Curve (AUC). This diverse array of metrics provides a thorough evaluation of each detector's prediction results, offering insights into different aspects of their performance.

Let $\theta$ be the score threshold at which the detector classifies a voice as genuine. The $FRR(\theta)$ and the $FAR(\theta)$ of the detector at the threshold $\theta$ are defined as follows:
\begin{align}
FAR(\theta) &= \frac{\sum\{fake\ voices\ with\ score > \theta\}}{\sum\{fake\ voices\}} \label{eq:pfa} \\
FRR(\theta) &= \frac{\sum\{real\ voices\ with\ score < \theta\}}{\sum\{genuine\ voices\}} \label{eq:pmiss}
\end{align}

\autoref{eq:pfa} and \autoref{eq:pmiss} are respectively decreasing and increasing functions of the threshold $\theta$. The EER is defined in \autoref{eq:eer} as the error rate at the specific threshold $\theta_{eer}$, where $FAR(\theta_{eer})$ and $FRR(\theta_{eer})$ are equal.
\begin{align}
EER &= FAR(\theta_{eer}) = FRR(\theta_{eer}) \label{eq:eer} \\
TAR(\theta) &= \frac{\sum\{real\ voices\ with\ score > \theta\}}{\sum\{genuine\ voices\}} \label{eq:tpr}
\end{align}

The AUC represents the area under the Receiver Operating Characteristic (ROC) curve. The ROC curve describes the trade-off between the False Acceptance Rate (FAR) and the True Acceptance Rate (TAR) (defined in \autoref{eq:tpr}) of a detector at various decision thresholds ($\theta$). AUC values range from 0 to 1, with higher values indicating better detector performance. An AUC close to 0.5 suggests the detector's performance is comparable to random guessing. ACC describes the detector's correct prediction rate on the dataset, as defined in \autoref{eq:acc}.
The F1-Score reflects the detector's sensitivity to FAR and FRR. Together with ACC, it provides a more comprehensive analysis of detector performance on manipulation subsets (\autoref{sec:eva}). The definition of F1-score is given in \autoref{eq:f1}. 
\begin{align}
ACC(\theta) &= \frac{\sum \{voices\ with\ correct\ score\}}{\sum \{ \text{fake voices} \} + \sum \{ \text{real voices} \}} \label{eq:acc} \\
F1\text{-}Score(\theta) &= \frac{(1-FAR(\theta)) + (1-FRR(\theta))}{2 \cdot (1-FAR(\theta)) \cdot (1-FRR(\theta))} \label{eq:f1}
\end{align}

In the subsequent sections of this paper, all discussed metrics are under the condition of $\theta = \theta_{eer}$.

\begin{table*}[h]
\caption{The EER (\%) values of all detectors, the difference between EER values on the English and Chinese datasets, and their FARs (\%), FRRs (\%) at the equal error point on the replay subset, as well as the difference between these two values. The numerical values appear in the following order: EER on the English dataset, EER on the Chinese dataset, the difference between these two, FAR on the English replay subdataset, FRR on the English replay subset, the difference between FAR and FRR on the English replay subset, FAR on the Chinese replay subset, FRR on the Chinese replay subset, and the difference between FAR and FRR on the Chinese replay subset.}
\label{tab:eer}
\centering
\scriptsize

\setlength{\tabcolsep}{1pt}

\begin{tabular}{>{\centering\arraybackslash}p{2.43cm}  >{\centering\arraybackslash}p{2.43cm}  >{\centering\arraybackslash}p{2.43cm}  >{\centering\arraybackslash}p{2.43cm}  >{\centering\arraybackslash}p{2.43cm}  >{\centering\arraybackslash}p{2.43cm}  >{\centering\arraybackslash}p{2.43cm}}
\toprule
% \rowcolor[gray]{1}
 & \multicolumn{1}{|>{\centering\arraybackslash}p{2.43cm}}{\textbf{AASIST~\cite{aasist}}} & \multicolumn{1}{|>{\centering\arraybackslash}p{2.43cm}}{\textbf{RawNet2~\cite{tak2021end}}} & \multicolumn{1}{|>{\centering\arraybackslash}p{2.43cm}}{\textbf{RawBoost~\cite{tak2022rawboost}}} & \multicolumn{1}{|>{\centering\arraybackslash}p{2.43cm}}{\textbf{OC-Softmax~\cite{Zhang_2021}}} & \multicolumn{1}{|>{\centering\arraybackslash}p{2.43cm}}{\textbf{RawGAT-ST~\cite{rawgatst}}} & \multicolumn{1}{|>{\centering\arraybackslash}p{2.43cm}}{\textbf{CLAD~\cite{wu2024clad}}} \\
\midrule
EN-EER & \multicolumn{1}{|>{\centering\arraybackslash}p{2.43cm}}{28.16} & \multicolumn{1}{|>{\centering\arraybackslash}p{2.43cm}}{25.22} & \multicolumn{1}{|>{\centering\arraybackslash}p{2.43cm}}{23.48} & \multicolumn{1}{|>{\centering\arraybackslash}p{2.43cm}}{32.37} & \multicolumn{1}{|>{\centering\arraybackslash}p{2.43cm}}{25.63} & \multicolumn{1}{|>{\centering\arraybackslash}p{2.43cm}}{20.00}  \\
\midrule
CN-EER & \multicolumn{1}{|>{\centering\arraybackslash}p{2.43cm}}{\cellcolor{gray!20}\textbf{51.88}} & \multicolumn{1}{|>{\centering\arraybackslash}p{2.43cm}}{33.23} & \multicolumn{1}{|>{\centering\arraybackslash}p{2.43cm}}{26.55} & \multicolumn{1}{|>{\centering\arraybackslash}p{2.43cm}}{28.86} & \multicolumn{1}{|>{\centering\arraybackslash}p{2.43cm}}{35.49} & \multicolumn{1}{|>{\centering\arraybackslash}p{2.43cm}}{33.05}  \\
\midrule
Difference & \multicolumn{1}{|>{\centering\arraybackslash}p{2.43cm}}{\cellcolor{gray!40}\textbf{-23.69}} & \multicolumn{1}{|>{\centering\arraybackslash}p{2.43cm}}{-8.01} & \multicolumn{1}{|>{\centering\arraybackslash}p{2.43cm}}{-3.07} & \multicolumn{1}{|>{\centering\arraybackslash}p{2.43cm}}{\cellcolor{gray!40}\textbf{+3.51}} & \multicolumn{1}{|>{\centering\arraybackslash}p{2.43cm}}{-9.87} & \multicolumn{1}{|>{\centering\arraybackslash}p{2.43cm}}{-13.05}  \\
\bottomrule
\\
\toprule
EN-FAR & \multicolumn{1}{|>{\centering\arraybackslash}p{2.43cm}}{17.88} & \multicolumn{1}{|>{\centering\arraybackslash}p{2.43cm}}{80.74} & \multicolumn{1}{|>{\centering\arraybackslash}p{2.43cm}}{47.24} & \multicolumn{1}{|>{\centering\arraybackslash}p{2.43cm}}{99.47} & \multicolumn{1}{|>{\centering\arraybackslash}p{2.43cm}}{15.25} & \multicolumn{1}{|>{\centering\arraybackslash}p{2.43cm}}{44.24}  \\
\midrule
EN-FRR & \multicolumn{1}{|>{\centering\arraybackslash}p{2.43cm}}{55.34} & \multicolumn{1}{|>{\centering\arraybackslash}p{2.43cm}}{12.65} & \multicolumn{1}{|>{\centering\arraybackslash}p{2.43cm}}{32.63} & \multicolumn{1}{|>{\centering\arraybackslash}p{2.43cm}}{0.00} & \multicolumn{1}{|>{\centering\arraybackslash}p{2.43cm}}{68.28} & \multicolumn{1}{|>{\centering\arraybackslash}p{2.43cm}}{17.99}  \\
\midrule
Difference & \multicolumn{1}{|>{\centering\arraybackslash}p{2.43cm}}{-37.45} & \multicolumn{1}{|>{\centering\arraybackslash}p{2.43cm}}{+68.09} & \multicolumn{1}{|>{\centering\arraybackslash}p{2.43cm}}{+14.60} & \multicolumn{1}{|>{\centering\arraybackslash}p{2.43cm}}{+99.47} & \multicolumn{1}{|>{\centering\arraybackslash}p{2.43cm}}{-53.03} & \multicolumn{1}{|>{\centering\arraybackslash}p{2.43cm}}{+17.99}  \\
\bottomrule
\\
\toprule
CN-FAR & \multicolumn{1}{|>{\centering\arraybackslash}p{2.43cm}}{54.97} & \multicolumn{1}{|>{\centering\arraybackslash}p{2.43cm}}{93.95} & \multicolumn{1}{|>{\centering\arraybackslash}p{2.43cm}}{53.68} & \multicolumn{1}{|>{\centering\arraybackslash}p{2.43cm}}{93.97} & \multicolumn{1}{|>{\centering\arraybackslash}p{2.43cm}}{30.21} & \multicolumn{1}{|>{\centering\arraybackslash}p{2.43cm}}{77.61}  \\
\midrule
CN-FRR & \multicolumn{1}{|>{\centering\arraybackslash}p{2.43cm}}{24.13} & \multicolumn{1}{|>{\centering\arraybackslash}p{2.43cm}}{2.39} & \multicolumn{1}{|>{\centering\arraybackslash}p{2.43cm}}{18.37} & \multicolumn{1}{|>{\centering\arraybackslash}p{2.43cm}}{1.34} & \multicolumn{1}{|>{\centering\arraybackslash}p{2.43cm}}{50.29} & \multicolumn{1}{|>{\centering\arraybackslash}p{2.43cm}}{10.97}  \\
\midrule
Difference & \multicolumn{1}{|>{\centering\arraybackslash}p{2.43cm}}{+30.84} & \multicolumn{1}{|>{\centering\arraybackslash}p{2.43cm}}{+91.55} & \multicolumn{1}{|>{\centering\arraybackslash}p{2.43cm}}{+35.32} & \multicolumn{1}{|>{\centering\arraybackslash}p{2.43cm}}{+92.63} & \multicolumn{1}{|>{\centering\arraybackslash}p{2.43cm}}{-20.08} & \multicolumn{1}{|>{\centering\arraybackslash}p{2.43cm}}{+66.63}  \\
\bottomrule
\\
\toprule
 & \multicolumn{1}{|>{\centering\arraybackslash}p{2.43cm}}{\textbf{Res-TSSDNet~\cite{hua2021tssdnet}}} & \multicolumn{1}{|>{\centering\arraybackslash}p{2.43cm}}{\textbf{RawNet2-Vocoder~\cite{sun2023rawnet2vocoder}}} & \multicolumn{1}{|>{\centering\arraybackslash}p{2.43cm}}{\textbf{AASIST2~\cite{tak2022aasist2}}} & \multicolumn{1}{|>{\centering\arraybackslash}p{2.43cm}}{\textbf{Raw PC-DARTS~\cite{ge2021rawpcdrats}}} & \multicolumn{1}{|>{\centering\arraybackslash}p{2.43cm}}{\textbf{RawBMamba~\cite{chen2024rawbmamba}}} & \multicolumn{1}{|>{\centering\arraybackslash}p{2.43cm}}{\textbf{SAMO~\cite{chen2024rawbmamba}}} \\
\midrule
EN-EER & \multicolumn{1}{|>{\centering\arraybackslash}p{2.43cm}}{\cellcolor{gray!20}\textbf{50.01}} & \multicolumn{1}{|>{\centering\arraybackslash}p{2.43cm}}{27.54} & \multicolumn{1}{|>{\centering\arraybackslash}p{2.43cm}}{\cellcolor{gray!60}\textbf{13.50}} & \multicolumn{1}{|>{\centering\arraybackslash}p{2.43cm}}{27.93} & \multicolumn{1}{|>{\centering\arraybackslash}p{2.43cm}}{27.43} & \multicolumn{1}{|>{\centering\arraybackslash}p{2.43cm}}{25.57}  \\
\midrule
CN-EER & \multicolumn{1}{|>{\centering\arraybackslash}p{2.43cm}}{49.91} & \multicolumn{1}{|>{\centering\arraybackslash}p{2.43cm}}{37.01} & \multicolumn{1}{|>{\centering\arraybackslash}p{2.43cm}}{\cellcolor{gray!60}\textbf{13.54}} & \multicolumn{1}{|>{\centering\arraybackslash}p{2.43cm}}{29.99} & \multicolumn{1}{|>{\centering\arraybackslash}p{2.43cm}}{32.48} & \multicolumn{1}{|>{\centering\arraybackslash}p{2.43cm}}{48.72}  \\
\midrule
Difference & \multicolumn{1}{|>{\centering\arraybackslash}p{2.43cm}}{+0.15} & \multicolumn{1}{|>{\centering\arraybackslash}p{2.43cm}}{-9.47} & \multicolumn{1}{|>{\centering\arraybackslash}p{2.43cm}}{\cellcolor{gray!40}-0.04} & \multicolumn{1}{|>{\centering\arraybackslash}p{2.43cm}}{-2.07} & \multicolumn{1}{|>{\centering\arraybackslash}p{2.43cm}}{-5.02} & \multicolumn{1}{|>{\centering\arraybackslash}p{2.43cm}}{-23.15}  \\
\bottomrule
\\
\toprule
EN-FAR & \multicolumn{1}{|>{\centering\arraybackslash}p{2.43cm}}{91.26} & \multicolumn{1}{|>{\centering\arraybackslash}p{2.43cm}}{61.10} & \multicolumn{1}{|>{\centering\arraybackslash}p{2.43cm}}{5.62} & \multicolumn{1}{|>{\centering\arraybackslash}p{2.43cm}}{99.65} & \multicolumn{1}{|>{\centering\arraybackslash}p{2.43cm}}{93.51} & \multicolumn{1}{|>{\centering\arraybackslash}p{2.43cm}}{29.71}  \\
\midrule
EN-FRR & \multicolumn{1}{|>{\centering\arraybackslash}p{2.43cm}}{9.03} & \multicolumn{1}{|>{\centering\arraybackslash}p{2.43cm}}{32.74} & \multicolumn{1}{|>{\centering\arraybackslash}p{2.43cm}}{72.59} & \multicolumn{1}{|>{\centering\arraybackslash}p{2.43cm}}{0.00} & \multicolumn{1}{|>{\centering\arraybackslash}p{2.43cm}}{1.37} & \multicolumn{1}{|>{\centering\arraybackslash}p{2.43cm}}{26.25}  \\
\midrule
Difference & \multicolumn{1}{|>{\centering\arraybackslash}p{2.43cm}}{+82.23} & \multicolumn{1}{|>{\centering\arraybackslash}p{2.43cm}}{+28.37} & \multicolumn{1}{|>{\centering\arraybackslash}p{2.43cm}}{-66.97} & \multicolumn{1}{|>{\centering\arraybackslash}p{2.43cm}}{+99.64} & \multicolumn{1}{|>{\centering\arraybackslash}p{2.43cm}}{+92.15} & \multicolumn{1}{|>{\centering\arraybackslash}p{2.43cm}}{-14.72}  \\
\bottomrule
\\
\toprule
CN-FAR & \multicolumn{1}{|>{\centering\arraybackslash}p{2.43cm}}{49.58} & \multicolumn{1}{|>{\centering\arraybackslash}p{2.43cm}}{85.84} & \multicolumn{1}{|>{\centering\arraybackslash}p{2.43cm}}{13.39} & \multicolumn{1}{|>{\centering\arraybackslash}p{2.43cm}}{99.53} & \multicolumn{1}{|>{\centering\arraybackslash}p{2.43cm}}{94.74} & \multicolumn{1}{|>{\centering\arraybackslash}p{2.43cm}}{46.13}  \\
\midrule
CN-FRR & \multicolumn{1}{|>{\centering\arraybackslash}p{2.43cm}}{49.97} & \multicolumn{1}{|>{\centering\arraybackslash}p{2.43cm}}{9.74} & \multicolumn{1}{|>{\centering\arraybackslash}p{2.43cm}}{33.79} & \multicolumn{1}{|>{\centering\arraybackslash}p{2.43cm}}{0.05} & \multicolumn{1}{|>{\centering\arraybackslash}p{2.43cm}}{4.84} & \multicolumn{1}{|>{\centering\arraybackslash}p{2.43cm}}{37.13}  \\
\midrule
Difference & \multicolumn{1}{|>{\centering\arraybackslash}p{2.43cm}}{-0.39} & \multicolumn{1}{|>{\centering\arraybackslash}p{2.43cm}}{+76.10} & \multicolumn{1}{|>{\centering\arraybackslash}p{2.43cm}}{-20.39} & \multicolumn{1}{|>{\centering\arraybackslash}p{2.43cm}}{+99.47} & \multicolumn{1}{|>{\centering\arraybackslash}p{2.43cm}}{+89.89} & \multicolumn{1}{|>{\centering\arraybackslash}p{2.43cm}}{+9.00}  \\
\bottomrule
\end{tabular}
\end{table*}

\section{Evaluation and Analysis}
\label{sec:eva}
\subsection{Evaluated Detectors}
\label{sec:evaluatedmethods}

We collected 12 open-source detectors that have gained significant attention in deepfake voice detection or demonstrated excellent performance in their original publications. Our focus was primarily on end-to-end detectors rather than traditional pipeline detectors, as the latter often require complex and time-intensive feature extraction processes for large-scale datasets. The selected detectors include \texttt{AASIST}~\cite{aasist} and \texttt{RawNet2}~\cite{tak2021end}, used as baselines in well-known challenges~\cite{2021asvspoof,2022add,jung2022sasv,yi2023add}, along with \texttt{RawBoost}~\cite{tak2022rawboost}, \texttt{OC-Softmax}~\cite{Zhang_2021}, \texttt{RawGAT-ST}~\cite{rawgatst}, \texttt{SAMO}~\cite{ding2023samo}, \texttt{Res-TSSDNet}~\cite{hua2021tssdnet}, \texttt{RawNet2-Vocoder}~\cite{sun2023rawnet2vocoder}, \texttt{AASIST2}~\cite{tak2022aasist2}, \texttt{Raw PC-DARTS}~\cite{ge2021rawpcdrats} and the latest detectors \texttt{RawBMamba}~\cite{chen2024rawbmamba} and \texttt{CLAD}~\cite{wu2024clad}.  We evaluate all detectors on our dataset as described in \autoref{sec:evamethods}.

\subsection{Evaluation Results}
\label{sec:aucandeer}

We examine the overall performance of various detectors. \autoref{tab:eer} shows the EER values for all detectors on our dataset. On the English dataset, EERs range from 13.50\% to 50.01\%, and on the Chinese dataset, from 13.54\% to 51.88\%. The best-performing detector is \texttt{AASIST2}, with the lowest EER on both datasets: 13.50\% for English and 13.54\% for Chinese, while all other detectors have EERs above 20\%. The worst performer on the English dataset is \texttt{Res-TSSDNet} (50.01\% EER), while on the Chinese dataset, \texttt{AASIST} has the highest EER at 51.88\%.
Most detectors perform better on the English dataset than on the Chinese dataset, likely due to their training on English data. EER differences between datasets indicate varying adaptability in cross-lingual detection. \texttt{AASIST2} has the smallest EER difference (-0.04\%) between the English and Chinese datasets, while \texttt{AASIST} shows the largest (-23.69\%). \texttt{SAMO} also has a significant gap, with its English EER 23.15 percentage points lower than its Chinese EER, second only to \texttt{AASIST}. Interestingly, \texttt{OC-Softmax} performs better on the Chinese dataset, with its EER 3.51 percentage points lower. \texttt{Res-TSSDNet} has EER values close to 50\% on both datasets, indicating poor performance across languages.
On the English dataset, aside from \texttt{AASIST2} (best) and \texttt{Res-TSSDNet} (worst), most detectors have similar EERs around 25\%. However, on the Chinese dataset, performance differences are more pronounced, highlighting variations in cross-lingual detection capabilities among detectors.

\begin{figure*}[htbp]
    \centering

    \begin{subfigure}{0.32\linewidth}
        \centering
        \includegraphics[width=\linewidth]{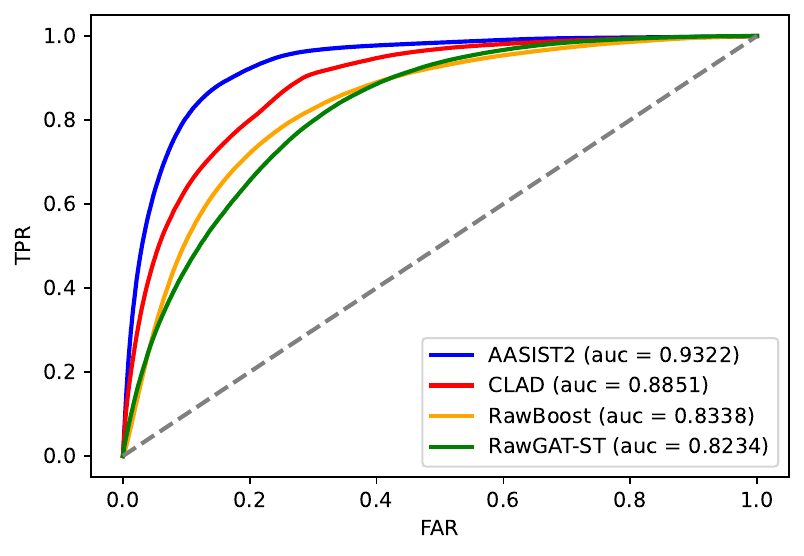}
        \caption{Top 1-4 AUC scores on English dataset.}
        \label{fig:enroc1}
    \end{subfigure}
    \hfill
    \begin{subfigure}{0.32\linewidth}
        \centering
        \includegraphics[width=\linewidth]{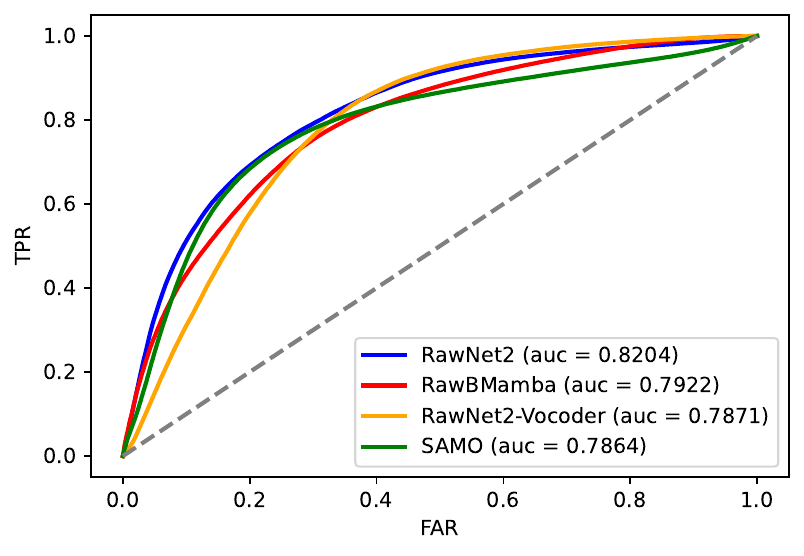}
        \caption{Top 5-8 AUC scores on English dataset.}
        \label{fig:enroc2}
    \end{subfigure}
    \hfill
    \begin{subfigure}{0.32\linewidth}
        \centering
        \includegraphics[width=\linewidth]{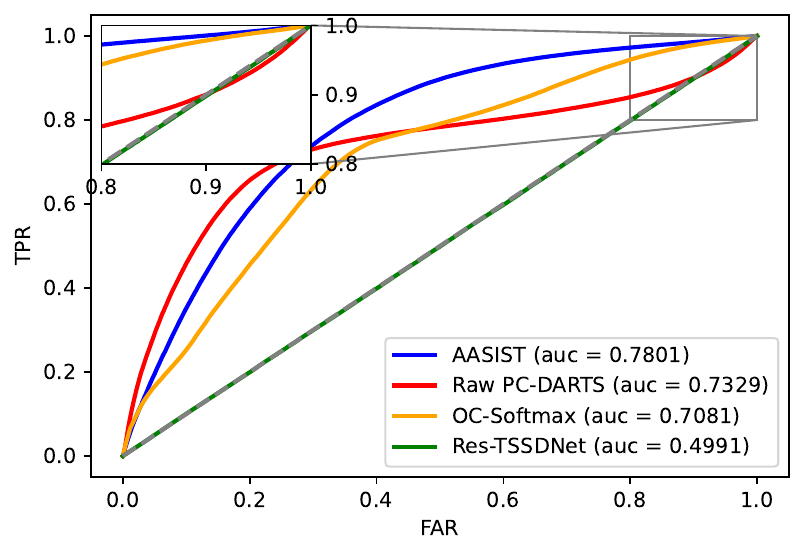}
        \caption{Top 9-12 AUC scores on English dataset.}
        \label{fig:enroc3}
    \end{subfigure}

    \begin{subfigure}{0.32\linewidth}
        \centering
        \includegraphics[width=\linewidth]{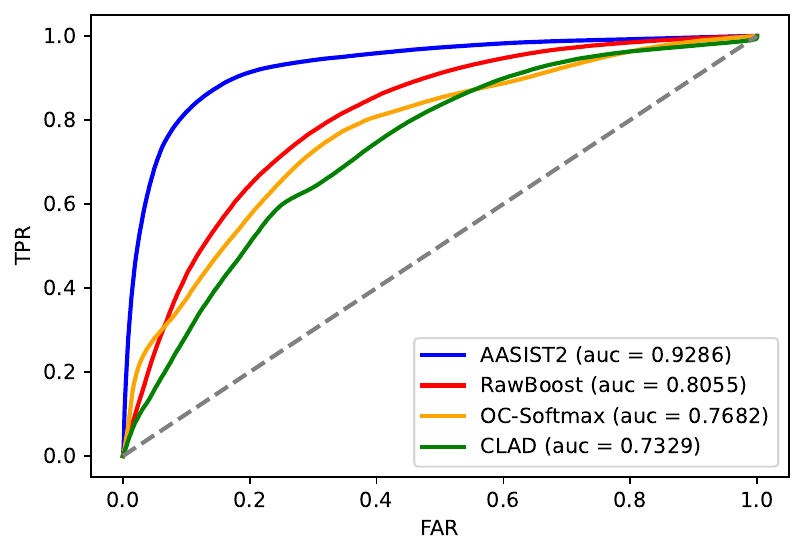}
        \caption{Top 1-4 AUC scores on Chinese dataset.}
        \label{fig:zhroc1}
    \end{subfigure}
    \hfill
    \begin{subfigure}{0.32\linewidth}
        \centering
        \includegraphics[width=\linewidth]{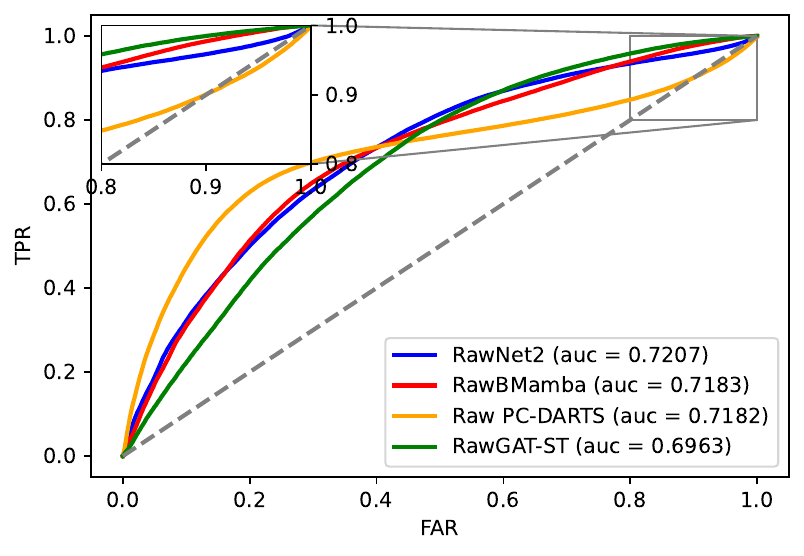}
        \caption{Top 5-8 AUC scores on Chinese dataset.}
        \label{fig:zhroc2}
    \end{subfigure}
    \hfill
    \begin{subfigure}{0.32\linewidth}
        \centering
        \includegraphics[width=\linewidth]{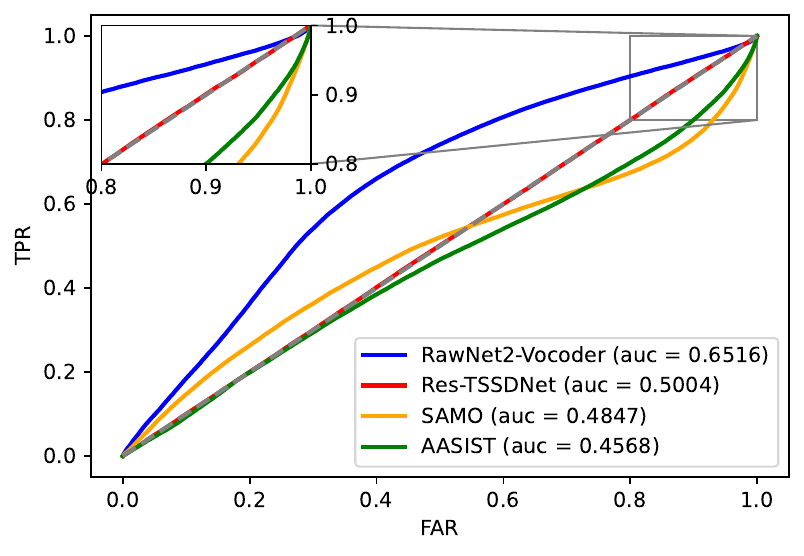}
        \caption{Top 9-12 AUC scores on Chinese dataset.}
        \label{fig:zhroc3}
    \end{subfigure}

    \caption{ROC curves for all detectors. (a), (b), and (c) show the performance of various detectors on the English dataset, while (d), (e), and (f) show their performance on the Chinese dataset.}
    \label{fig:roc}
\end{figure*}

We further assess each detector's robustness using AUC scores across different discrimination thresholds. \autoref{fig:roc} presents the AUC scores and ROC curves for each detector's predictions. As shown in \autoref{fig:enroc1} and \autoref{fig:zhroc1}, only \texttt{AASIST2} achieved AUC scores above 0.9 on both language datasets. On the English dataset, four detectors (\texttt{CLAD}, \texttt{RawBoost}, \texttt{RawGAT-ST}, and \texttt{RawNet2}) have AUC scores between 0.8 and 0.9, while all other detectors, except \texttt{Res-TSSDNet}, fall between 0.7 and 0.8.  On the Chinese dataset, only \texttt{RawBoost} has an AUC score above 0.8, while five other detectors (\texttt{OC-Softmax}, \texttt{CLAD}, \texttt{RawNet2}, \texttt{RawBmamba} and \texttt{RawNet2-Vocoder}) scoring between 0.7 and 0.8. 
Notably, \texttt{AASIST2}, \texttt{CLAD}, and \texttt{RawBoost} consistently rank in the top four for AUC scores in both languages.
\texttt{Res-TSSDNet} performs poorly on both datasets, with AUC scores close to 0.5 (\autoref{fig:enroc3} and \autoref{fig:zhroc3}). On the Chinese dataset, \texttt{SAMO} and \texttt{AASIST} also show similarly poor performance. The ROC curves for these detectors nearly overlap with the random prediction line, indicating extremely poor performance. \texttt{Res-TSSDNet} shows similar performance on both languages, which we discuss further in Appendix \autoref{sec:restssdnet}. Interestingly, \texttt{Raw PC-DARTS} exhibits reverse prediction behavior under certain thresholds on both datasets (\autoref{fig:enroc3} and \autoref{fig:zhroc2}), as its ROC curves cross the random prediction line. \autoref{fig:zhroc3} shows that \texttt{SAMO} and \texttt{AASIST} also display this phenomenon on the Chinese dataset.

Among the 12 evaluated detectors, \texttt{AASIST2} demonstrates the best overall performance and robustness. However, its EER on our dataset significantly underperforms compared to the 0.82\% reported in the original paper. 
Other detectors also show notable discrepancies from their originally reported performances, for instance, \texttt{RawBoost}'s best EER was 5.31\% in the original paper but reached 23.48\% in \bench{}. These findings raise concerns about the practical effectiveness of deepfake voice detection in real-world scenarios.

\subsection{Effect of Manipulations}
We analyze the detectors' performance across various manipulated subsets of our dataset, using each detector's performance on the standardized (std.) subset as a baseline. This approach helps identify specific differences under various manipulations and reveals potential optimization methods. \autoref{tab:accandf1en} and \autoref{tab:accandf1zh} in the Appendix present the ACCs and F1-Scores for all detectors at $\theta_{eer}$ on various subsets. Due to space constraints, we focus on the top four detectors by AUC score in both languages, while discussing the remaining detectors only for their unique performance characteristics.

Analysis of the std. subset performance reveals \texttt{AASIST2} as the standout performer, achieving ACCs above 90\% on both language datasets (\autoref{tab:accandf1en} and \autoref{tab:accandf1zh}). On the English dataset, most detectors exceed 80\% ACC, with \texttt{CLAD} peaking at 85.25\%. Only \texttt{RawNet2} and \texttt{AASIST} fall slightly below 80\%. The Chinese dataset shows a general performance decline, except for \texttt{OC-Softmax}. \texttt{AASIST} experiences the most severe drop to 48.72\% ACC. Notably, \texttt{CLAD}, despite ranking fourth in AUC score, achieves only 68.67\% ACC on the Chinese std. subset. These results align with the AUC performance evaluation in \autoref{sec:aucandeer}, highlighting varied cross-lingual detection capabilities.

\begin{figure*}[htbp]
    \begin{subfigure}{\linewidth}
        \centering
        \includegraphics[width=\linewidth]{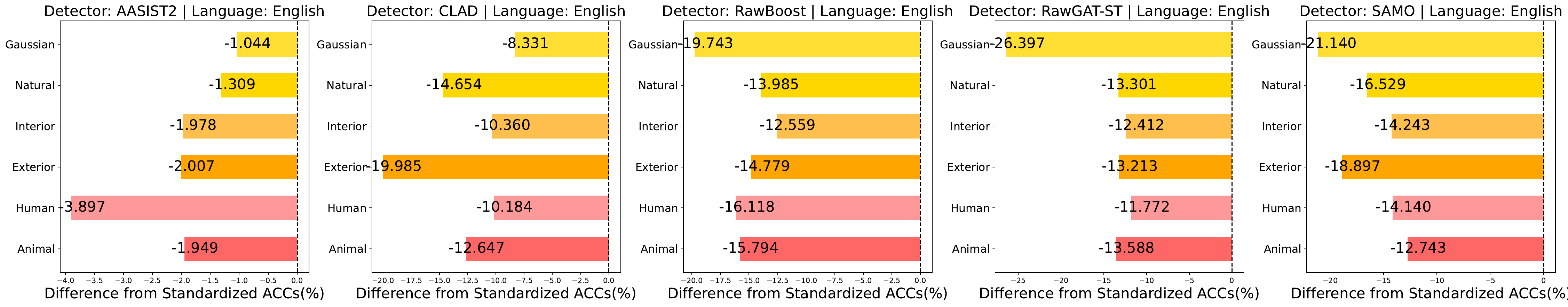}
        \caption{ACC differences for English dataset: the top four AUC-ranked detectors and SAMO on SNR 15dB noise-injected vs. std. subsets}
        \label{fig:en_noised_diff}
    \end{subfigure}

    % \vspace{1em} 
    \begin{subfigure}{\linewidth}
        \centering
        \includegraphics[width=\linewidth]{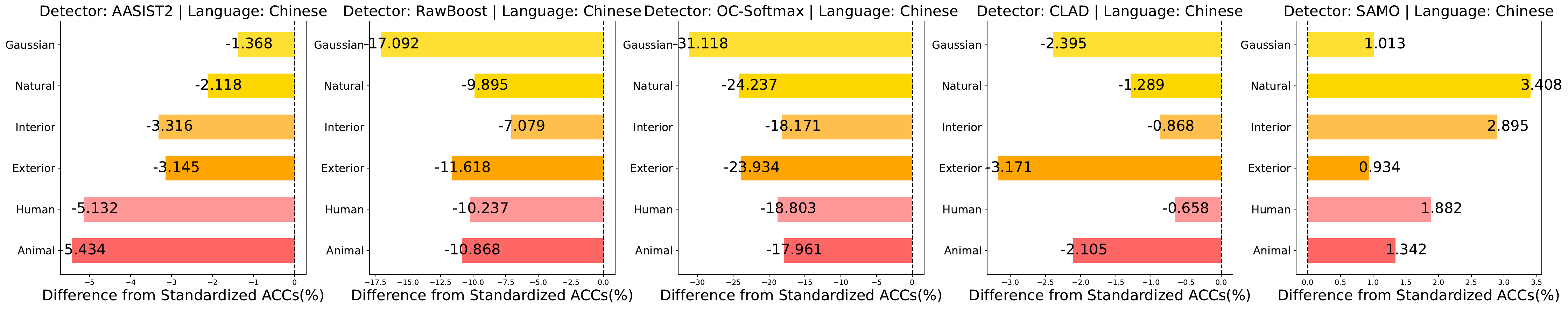}
        \caption{ACC differences for Chinese dataset: the top four AUC-ranked detectors and SAMO on SNR 15dB noise-injected vs. std. subsets}
        \label{fig:zh_noise_diff}
    \end{subfigure}
    \caption{ACC differences: the top four AUC-ranked detectors and SAMO on SNR 15dB noise-injected vs. std. subsets}
    \label{fig:noised_diff}
\end{figure*}

\noindent\textbf{Effect of Noise Injection.} \autoref{fig:noised_diff} illustrates ACC differences between 15 dB noise-injected and std. subsets for the top four detectors. All detectors show ACC declines across both datasets, with some interesting exceptions. \texttt{SAMO} experiences a significant drop on the English dataset but a slight increase on the Chinese dataset post-noise injection. Different noise types impact detectors variably. For instance, Human noise notably affects \texttt{AASIST}, while Gaussian noise has the least impact. Conversely, \texttt{RawBoost} is most affected by Gaussian noise and least by Interior noise.
The top four detectors show similar patterns of noise impact across languages. \texttt{AASIST2} is least affected by noise on the English dataset, while \texttt{CLAD} is most resilient on the Chinese dataset. 
Overall, \textbf{when facing noise injection, most detectors may experience a very significant performance decline}. Only \texttt{AASIST2} is the least affected, with a maximum ACC drop of 5.434\% on the Chinese dataset with Animal-type noise injection.
The impact varies across different noise types and detectors, underscoring the complexity of noise effects on deepfake voice detection.

The higher the SNR, the smaller the impact of noise on the voice, and the higher the quality of the voice. To explore the impact of different SNRs on the performance of the detectors, in the manually injected noise experiments, we set different SNRs of 15dB, 20dB, and 25dB for each type of noise.
\autoref{fig:noisesnr} in the Appendix shows the ACC of the top four AUC-ranked detectors under different SNRs. On the English dataset, for all the detectors evaluated in this work except \texttt{Res-TSSDNet}, the prediction ACC increases with the increase of SNR. On the Chinese dataset, except for \texttt{AASIST}, \texttt{Res-TSSDnet}, and \texttt{SAMO}, the ACCs of the remaining detectors also increases with the increase of the SNR, and only \texttt{RawBmamba} shows the opposite trend under Gaussian noise injection.
The impact of SNRs on different detectors varies. As shown in \autoref{fig:ennoise}, the ACC change of \texttt{AASIST2} is minimal under different SNRs, while \texttt{CLAD}, \texttt{RawBoost}, and \texttt{RawGat-ST} show more significant changes.
The difference in data language can also lead to different performance of the detectors under changes in SNRs. On the English dataset, \texttt{CLAD} shows a significant change in ACC, but on the Chinese dataset, this change is minimal.

\noindent\textbf{Effect of Volume Control.} 
For all detectors, the impact of volume control on detection performance shows no clear linear relationship based on the ACCs and F1-Scores results in \autoref{tab:accandf1en} and \autoref{tab:accandf1zh}. Generally, volume control mainly adjusts the amplitude of the voice without significantly affecting other important features, such as spectral characteristics and timbre. \autoref{fig:spectrogram_similarity} in the Appendix shows the cosine similarity of spectrograms calculated for the same voice at different volume levels, further confirming this point. Detectors typically rely on specific features to identify deepfake voices. During data collection, the range of volume adjustments was determined through human listening tests to ensure that the voice content remained audible. Therefore, the impact on features is even more limited, resulting in relatively stable performance across all detectors when facing volume control.

\begin{figure}[h]
    \centering
    \includegraphics[width=\linewidth]{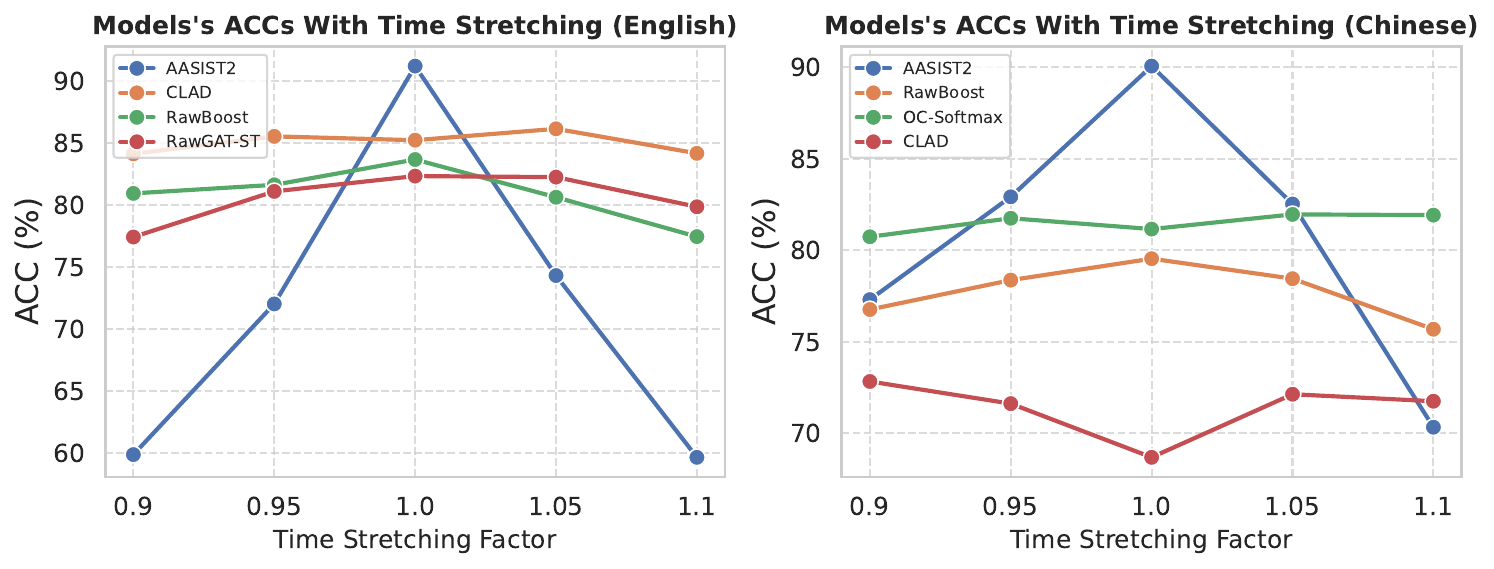}
    \caption{The ACC changes of the top four AUC-ranked detectors under different time stretching factors. \texttt{AASIST2} showed a performance that was strikingly different from previous evaluations. Its performance fluctuated severely under time stretching.}
    \label{fig:timestretch}
\end{figure}

\noindent\textbf{Effect of Time Stretching.} 
\autoref{fig:timestretch} shows the ACCs of various detectors under different factors of time stretching. \textbf{Although the \texttt{AASIST2} performs exceptionally well in noise injection scenarios, it exhibits the most significant performance degradation among the top four AUC-ranked detectors when faced with time-stretching manipulations.} Moreover, the detection performance of \texttt{AASIST2} worsens as the magnitude of time stretching increases. In the English dataset, the performance of the other three detectors remains relatively stable, with \texttt{CLAD} being the most consistent detector. Similarly, in the Chinese dataset, the three detectors other than \texttt{AASIST2} also demonstrate stable performance, with the \texttt{OC-Softmax} showing the best results. \texttt{AASIST2} relies on the frontend features of wav2vec 2.0~\cite{baevski2020wav2vec20frameworkselfsupervised}, which is sensitive to temporal information. Time-stretching manipulations can cause shifts or distortions in some key temporally related features. Consequently,  \texttt{AASIST2} has relatively weak resistance to time stretching interference, resulting in a noticeable decline in performance.

\noindent\textbf{Effect of Voice Resampling.} 
As shown in the \autoref{tab:accandf1en} and \autoref{tab:accandf1zh}, in terms of ACCs and F1-Scores, all detectors, except for \texttt{Raw PC-DARTS} and those detectors with near-random prediction AUC on their respective language datasets, experienced severe performance degradation after voice resampling. In the English dataset, \texttt{CLAD} and \texttt{RawNet2} saw their F1-Scores drop below 30\% (29.75\% and 25.85\% respectively) at a sample rate of 44100~Hz. In contrast, \texttt{Raw PC-DARTS} maintained ACCs and F1-Scores close to the std. subset for both English and Chinese datasets.
\textbf{When resampling from the std. subset to higher sampling rates, more feature information is lost in the same length of sampling sequence.} We speculate that this is the main reason for the performance decline of most detectors when facing resampling.
\texttt{Raw PC-DARTS} employs learnable \texttt{Sinc} filters, which suggests that it can still extract key features when confronted with different sample rates. This also explains its outstanding performance stability when dealing with resampled datasets.

\noindent\textbf{Effect of Replay.} 
As observed from the \autoref{tab:accandf1en} and \autoref{tab:accandf1zh}, all detectors show a significant decrease in ACCs and F1-Scores when faced with replay subset. We further analyzed this using the FARs and FRRs of each detector. \autoref{tab:eer} presents the FARs, FRRs, and the difference between them for each detector on the replay subset.
\texttt{RawNet2}, \texttt{Raw PC-DARTS}, \texttt{OC-Softmax}, and \texttt{RawBMamba} exhibit FAR values significantly higher than their FRR values in both languages. This indicates that these detectors are highly likely to classify replayed voices as real voices, demonstrating a clear lack of defense against easily implementable replay attacks.
\texttt{CLAD} shows a FAR 17.99 percentage points higher than its FRR in the English dataset, but this difference increases to 66.63 in the Chinese dataset, suggesting the detector's vulnerability to replay attacks.
In contrast, \texttt{AASIST2} and \texttt{RawGAT-ST} have FAR values noticeably lower than their FRR values in both languages. This implies that these two detectors tend to classify replayed voices as deepfake voices, making them more suitable for hindering deepfake voices that have undergone replay operations.

\noindent\textbf{Effect of Fade In \& Out.} 
Observing the statistical data in the fade-in and fade-out parts of the \autoref{tab:accandf1en} and \autoref{tab:accandf1zh}, we observed that on the English dataset, \texttt{AASIST2}, \texttt{CLAD}, and \texttt{RooBoost} show no significant changes in ACCs when facing linear and exponential fade-in and fade-out. The proportion of fade influences the ACC, but the impact is minimal. However, when facing logarithmic fade, the performance decline of these three detectors is more noticeable. From the changes in FARs and FRRs, we can infer that the detectors tend to classify deepfake voices as real voices. \texttt{RawGAT-ST} consistently performs the worst among the top four AUC-ranked detectors. Its ACC declines noticeably in all three shapes of fade, with the decline becoming more pronounced as the proportion increases. It also shows a more severe tendency to misclassify deepfake voices as real voices.
On the Chinese dataset, most detectors exhibited lower performance compared to their results on the English dataset, and the overall trend of changes is consistent with that of the English dataset. Interestingly, \texttt{OC-Softmax} shows surprisingly stable ACCs on the Chinese dataset, outperforming its performance on the English dataset.
In total, \texttt{AASIST2} demonstrates the best robustness when facing fade in \& out. Logarithmic fade has the most significant impact on detectors' performance. When the proportion of logarithmic fade increases to 0.3, some detectors like \texttt{RawGAT-ST} tend to largely classify deepfake voices as real voices, while linear fade has the least impact. Detectors show noticeable performance differences across datasets of different languages.

 \section{Models vs. Humans}

This section analyzes user study results, categorizing deepfake voices by deception levels (\autoref{sec:studydesign}). We assess each detector's performance across these levels. Additionally, given large language models (LLMs)' versatility in various tasks~\cite{hou2023large}, we explore an open-source multimodal LLM (MLLM)'s potential for deepfake voice identification.

\subsection{User Study Results and Deception Levels}
\label{sec:userstudyresult}

Observing the overall performance of participants across the sample dataset, we compiled the statistics from the three user study experimental rounds and their average results, as shown in \autoref{tab:userstudyresult}. For English, participants achieved average ACC of 56.69\% and F1-Score of 60.61\%. For Chinese, average ACC was 77.94\% with F1-Score of 79.33\%. Results were consistent across rounds, indicating stable human discernment of deepfake voices.
Therefore, we can use the actual deception results of deepfake voices on users from the user study to grade the deepfake voices.
Using the method in \autoref{sec:userstudy}, we created voice combinations and classified them into deception levels 0-2 based on human deception success. For English, 1,326 combinations were produced (210 level 0, 829 level 1, 287 level 2). For Chinese, 741 combinations were created (440 level 0, 184 level 1, 117 level 2). \autoref{tab:humanmodel} displays participants' FARs for these three deception levels.

\begin{table}[h]
\centering
% \scriptsize
\footnotesize
\renewcommand{\arraystretch}{1.5}
\caption{The results of the three user study experimental rounds. Results are consistent in both languages, indicating that participants' judgment abilities remain stable.}
\label{tab:userstudyresult}
\resizebox{0.75\linewidth}{!}{%
\begin{tabular}{|c|c|c|c|c|}
\hline
\multirow{2}{*}{\textbf{No.}} & \multicolumn{2}{c|}{\textbf{English}} & \multicolumn{2}{c|}{\textbf{Chinese}} \\
\cline{2-5}
 & \textbf{ACC} & \textbf{F1-Score} & \textbf{ACC} & \textbf{F1-Score} \\
\hline
 \bf  Exp.R1  &  59.28 & 62.30 & 78.20 & 80.06  \\
\hline
 \bf  Exp.R2  &  55.90 & 59.51 & 78.10 & 78.94  \\
\hline
 \bf Exp.R3  &  54.92 & 60.01 & 77.50 &  79.01  \\
\hline
 \bf Average  &  56.69 & 60.61 & 77.94 & 79.33 \\
\hline
\end{tabular}%
}
\end{table}

\begin{table*}[h]
\caption{FARs (\%) of humans and various detectors on different levels of deepfake voices. Rows 2-4 and 9-11 of the table show the results on the English dataset, while the other rows present the results on the Chinese dataset.}

\centering
\scriptsize

\setlength{\tabcolsep}{1pt}

\begin{tabular}{>{\centering\arraybackslash}p{2.125cm}  >{\centering\arraybackslash}p{2.125cm}  >{\centering\arraybackslash}p{2.125cm}  >{\centering\arraybackslash}p{2.125cm}  >{\centering\arraybackslash}p{2.125cm}  >{\centering\arraybackslash}p{2.125cm}  >{\centering\arraybackslash}p{2.125cm} >
{\centering\arraybackslash}p{2.125cm} }
\toprule
 & \multicolumn{1}{|>{\centering\arraybackslash}p{2.125cm}}{\textbf{Human}}& \multicolumn{1}{|>{\centering\arraybackslash}p{2.125cm}}{\textbf{AASIST}} & \multicolumn{1}{|>{\centering\arraybackslash}p{2.125cm}}{\textbf{RawNet2}} & \multicolumn{1}{|>{\centering\arraybackslash}p{2.125cm}}{\textbf{RawBoost}} & \multicolumn{1}{|>{\centering\arraybackslash}p{2.125cm}}{\textbf{OC-Softmax}} & \multicolumn{1}{|>{\centering\arraybackslash}p{2.125cm}}{\textbf{RawGAT-ST}} & \multicolumn{1}{|>{\centering\arraybackslash}p{2.125cm}}{\textbf{CLAD}}\\
\midrule
EN-Level0 & \multicolumn{1}{|>{\centering\arraybackslash}p{2.125cm}}{18.97} & \multicolumn{1}{|>{\centering\arraybackslash}p{2.125cm}}{\cellcolor[rgb]{1,0.9,0.9} 28.33} & \multicolumn{1}{|>{\centering\arraybackslash}p{2.125cm}}{\cellcolor[rgb]{1,0.9,0.9} 25.71} & \multicolumn{1}{|>{\centering\arraybackslash}p{2.125cm}}{\cellcolor[rgb]{1,0.9,0.9} 23.81} & \multicolumn{1}{|>{\centering\arraybackslash}p{2.125cm}}{\cellcolor[rgb]{1,0.9,0.9} 30.48} & \multicolumn{1}{|>{\centering\arraybackslash}p{2.125cm}}{\cellcolor[rgb]{1,0.9,0.9} 24.04} & \multicolumn{1}{|>{\centering\arraybackslash}p{2.125cm}}{\cellcolor[rgb]{0.9,1,0.9} 17.62}  \\
\midrule
EN-Level1 & \multicolumn{1}{|>{\centering\arraybackslash}p{2.125cm}}{51.67}& \multicolumn{1}{|>{\centering\arraybackslash}p{2.125cm}}{\cellcolor[rgb]{0.9,1,0.9}28.65} & \multicolumn{1}{|>{\centering\arraybackslash}p{2.125cm}}{\cellcolor[rgb]{0.9,1,0.9}26.48} & \multicolumn{1}{|>{\centering\arraybackslash}p{2.125cm}}{\cellcolor[rgb]{0.9,1,0.9}26.54} & \multicolumn{1}{|>{\centering\arraybackslash}p{2.125cm}}{\cellcolor[rgb]{0.9,1,0.9}32.31} & \multicolumn{1}{|>{\centering\arraybackslash}p{2.125cm}}{\cellcolor[rgb]{0.9,1,0.9}26.90} & \multicolumn{1}{|>{\centering\arraybackslash}p{2.125cm}}{\cellcolor[rgb]{0.9,1,0.9} 22.56}  \\
\midrule
 EN-Level2 &  \multicolumn{1}{|>{\centering\arraybackslash}p{2.125cm}}{ 82.64} & \multicolumn{1}{|>{\centering\arraybackslash}p{2.125cm}}{\cellcolor[rgb]{0.9,1,0.9}30.14} & \multicolumn{1}{|>{\centering\arraybackslash}p{2.125cm}}{\cellcolor[rgb]{0.9,1,0.9}29.27} & \multicolumn{1}{|>{\centering\arraybackslash}p{2.125cm}}{\cellcolor[rgb]{0.9,1,0.9} 24.74} & \multicolumn{1}{|>{\centering\arraybackslash}p{2.125cm}}{\cellcolor[rgb]{0.9,1,0.9}35.54} & \multicolumn{1}{|>{\centering\arraybackslash}p{2.125cm}}{\cellcolor[rgb]{0.9,1,0.9}26.13} & \multicolumn{1}{|>{\centering\arraybackslash}p{2.125cm}}{\cellcolor[rgb]{0.9,1,0.9}24.39}  \\
\midrule
CN-Level0 &  \multicolumn{1}{|>{\centering\arraybackslash}p{2.125cm}}{ 4.20} & \multicolumn{1}{|>{\centering\arraybackslash}p{2.125cm}}{\cellcolor[rgb]{1,0.9,0.9} 57.84} & \multicolumn{1}{|>{\centering\arraybackslash}p{2.125cm}}{\cellcolor[rgb]{1,0.9,0.9} 29.20} & \multicolumn{1}{|>{\centering\arraybackslash}p{2.125cm}}{\cellcolor[rgb]{1,0.9,0.9} 14.20} & \multicolumn{1}{|>{\centering\arraybackslash}p{2.125cm}}{\cellcolor[rgb]{1,0.9,0.9} 22.72} & \multicolumn{1}{|>{\centering\arraybackslash}p{2.125cm}}{\cellcolor[rgb]{1,0.9,0.9} 33.18} & \multicolumn{1}{|>{\centering\arraybackslash}p{2.125cm}}{\cellcolor[rgb]{1,0.9,0.9} 33.75}  \\
\midrule
CN-Level1 &  \multicolumn{1}{|>{\centering\arraybackslash}p{2.125cm}}{50.54} & \multicolumn{1}{|>{\centering\arraybackslash}p{2.125cm}}{\cellcolor[rgb]{0.9,1,0.9}47.01} & \multicolumn{1}{|>{\centering\arraybackslash}p{2.125cm}}{\cellcolor[rgb]{0.9,1,0.9}46.20} & \multicolumn{1}{|>{\centering\arraybackslash}p{2.125cm}}{\cellcolor[rgb]{0.9,1,0.9}47.28} & \multicolumn{1}{|>{\centering\arraybackslash}p{2.125cm}}{\cellcolor[rgb]{0.9,1,0.9} 40.76} & \multicolumn{1}{|>{\centering\arraybackslash}p{2.125cm}}{\cellcolor[rgb]{0.9,1,0.9}47.28} & \multicolumn{1}{|>{\centering\arraybackslash}p{2.125cm}}{\cellcolor[rgb]{0.9,1,0.9}41.03}  \\
\midrule
CN-Level2 &  \multicolumn{1}{|>{\centering\arraybackslash}p{2.125cm}}{ 87.61} & \multicolumn{1}{|>{\centering\arraybackslash}p{2.125cm}}{\cellcolor[rgb]{0.9,1,0.9} 33.76} & \multicolumn{1}{|>{\centering\arraybackslash}p{2.125cm}}{\cellcolor[rgb]{0.9,1,0.9}48.29} & \multicolumn{1}{|>{\centering\arraybackslash}p{2.125cm}}{\cellcolor[rgb]{0.9,1,0.9}47.44} & \multicolumn{1}{|>{\centering\arraybackslash}p{2.125cm}}{\cellcolor[rgb]{0.9,1,0.9}48.72} & \multicolumn{1}{|>{\centering\arraybackslash}p{2.125cm}}{\cellcolor[rgb]{0.9,1,0.9}45.73} & \multicolumn{1}{|>{\centering\arraybackslash}p{2.125cm}}{\cellcolor[rgb]{0.9,1,0.9}34.62}  \\
\bottomrule
\\
\toprule
 & \multicolumn{1}{|>{\centering\arraybackslash}p{2.125cm}}{\textbf{Qwen-audio2}}& \multicolumn{1}{|>{\centering\arraybackslash}p{2.125cm}}{\textbf{Res-TSSDNet}} & \multicolumn{1}{|>{\centering\arraybackslash}p{2.125cm}}{\textbf{RawNet2-Vocoder}} & \multicolumn{1}{|>{\centering\arraybackslash}p{2.125cm}}{\textbf{AASIST2}} & \multicolumn{1}{|>{\centering\arraybackslash}p{2.125cm}}{\textbf{Raw PC-DARTS}} & \multicolumn{1}{|>{\centering\arraybackslash}p{2.125cm}}{\textbf{RawBmamba}} & \multicolumn{1}{|>{\centering\arraybackslash}p{2.125cm}}{\textbf{SAMO}} \\
\midrule
EN-Level0 & \multicolumn{1}{|>{\centering\arraybackslash}p{2.125cm}}{31.67} & \multicolumn{1}{|>{\centering\arraybackslash}p{2.125cm}}{\cellcolor[rgb]{1,0.9,0.9} 42.86} & \multicolumn{1}{|>{\centering\arraybackslash}p{2.125cm}}{\cellcolor[rgb]{1,0.9,0.9} 26.43} & \multicolumn{1}{|>{\centering\arraybackslash}p{2.125cm}}{\cellcolor[rgb]{0.9,1,0.9} 9.52} & \multicolumn{1}{|>{\centering\arraybackslash}p{2.125cm}}{\cellcolor[rgb]{1,0.9,0.9} 27.62} & \multicolumn{1}{|>{\centering\arraybackslash}p{2.125cm}}{\cellcolor[rgb]{1,0.9,0.9} 26.19} & \multicolumn{1}{|>{\centering\arraybackslash}p{2.125cm}}{\cellcolor[rgb]{1,0.9,0.9} 24.76}  \\
\midrule
EN-Level1 & \multicolumn{1}{|>{\centering\arraybackslash}p{2.125cm}}{32.15}& \multicolumn{1}{|>{\centering\arraybackslash}p{2.125cm}}{\cellcolor[rgb]{1,0.9,0.9} 51.69} & \multicolumn{1}{|>{\centering\arraybackslash}p{2.125cm}}{\cellcolor[rgb]{0.9,1,0.9}29.07} & \multicolumn{1}{|>{\centering\arraybackslash}p{2.125cm}}{\cellcolor[rgb]{0.9,1,0.9} 14.60} & \multicolumn{1}{|>{\centering\arraybackslash}p{2.125cm}}{\cellcolor[rgb]{0.9,1,0.9}28.59} & \multicolumn{1}{|>{\centering\arraybackslash}p{2.125cm}}{\cellcolor[rgb]{0.9,1,0.9}29.07} & \multicolumn{1}{|>{\centering\arraybackslash}p{2.125cm}}{\cellcolor[rgb]{0.9,1,0.9}27.20}  \\
\midrule
 EN-Level2 &  \multicolumn{1}{|>{\centering\arraybackslash}p{2.125cm}}{32.06} & \multicolumn{1}{|>{\centering\arraybackslash}p{2.125cm}}{\cellcolor[rgb]{0.9,1,0.9}54.36} & \multicolumn{1}{|>{\centering\arraybackslash}p{2.125cm}}{\cellcolor[rgb]{0.9,1,0.9}30.14} & \multicolumn{1}{|>{\centering\arraybackslash}p{2.125cm}}{\cellcolor[rgb]{0.9,1,0.9} 18.29} & \multicolumn{1}{|>{\centering\arraybackslash}p{2.125cm}}{\cellcolor[rgb]{0.9,1,0.9}32.75} & \multicolumn{1}{|>{\centering\arraybackslash}p{2.125cm}}{\cellcolor[rgb]{0.9,1,0.9}29.97} & \multicolumn{1}{|>{\centering\arraybackslash}p{2.125cm}}{\cellcolor[rgb]{0.9,1,0.9}28.92}  \\
\midrule
CN-Level0 &  \multicolumn{1}{|>{\centering\arraybackslash}p{2.125cm}}{34.87} & \multicolumn{1}{|>{\centering\arraybackslash}p{2.125cm}}{\cellcolor[rgb]{1,0.9,0.9} 48.30} & \multicolumn{1}{|>{\centering\arraybackslash}p{2.125cm}}{\cellcolor[rgb]{1,0.9,0.9} 29.32} & \multicolumn{1}{|>{\centering\arraybackslash}p{2.125cm}}{\cellcolor[rgb]{1,0.9,0.9}  9.55} & \multicolumn{1}{|>{\centering\arraybackslash}p{2.125cm}}{\cellcolor[rgb]{1,0.9,0.9} 21.82} & \multicolumn{1}{|>{\centering\arraybackslash}p{2.125cm}}{\cellcolor[rgb]{1,0.9,0.9} 30.91} & \multicolumn{1}{|>{\centering\arraybackslash}p{2.125cm}}{\cellcolor[rgb]{1,0.9,0.9} 56.02}  \\
\midrule
CN-Level1 &  \multicolumn{1}{|>{\centering\arraybackslash}p{2.125cm}}{28.80} & \multicolumn{1}{|>{\centering\arraybackslash}p{2.125cm}}{\cellcolor[rgb]{0.9,1,0.9} 48.91} & \multicolumn{1}{|>{\centering\arraybackslash}p{2.125cm}}{\cellcolor[rgb]{1,0.9,0.9} 50.82} & \multicolumn{1}{|>{\centering\arraybackslash}p{2.125cm}}{\cellcolor[rgb]{0.9,1,0.9} 27.17} & \multicolumn{1}{|>{\centering\arraybackslash}p{2.125cm}}{\cellcolor[rgb]{0.9,1,0.9} 44.84} & \multicolumn{1}{|>{\centering\arraybackslash}p{2.125cm}}{\cellcolor[rgb]{0.9,1,0.9} 44.02} & \multicolumn{1}{|>{\centering\arraybackslash}p{2.125cm}}{\cellcolor[rgb]{0.9,1,0.9} 45.92}  \\
\midrule
CN-Level2 &  \multicolumn{1}{|>{\centering\arraybackslash}p{2.125cm}}{ 28.63} & \multicolumn{1}{|>{\centering\arraybackslash}p{2.125cm}}{\cellcolor[rgb]{0.9,1,0.9} 45.30} & \multicolumn{1}{|>{\centering\arraybackslash}p{2.125cm}}{\cellcolor[rgb]{0.9,1,0.9} 48.29} & \multicolumn{1}{|>{\centering\arraybackslash}p{2.125cm}}{\cellcolor[rgb]{0.9,1,0.9} 25.64} & \multicolumn{1}{|>{\centering\arraybackslash}p{2.125cm}}{\cellcolor[rgb]{0.9,1,0.9}  49.57} & \multicolumn{1}{|>{\centering\arraybackslash}p{2.125cm}}{\cellcolor[rgb]{0.9,1,0.9} 35.90} & \multicolumn{1}{|>{\centering\arraybackslash}p{2.125cm}}{\cellcolor[rgb]{0.9,1,0.9} 33.33}  \\
\bottomrule
\label{tab:humanmodel}
\end{tabular}
\end{table*}

\subsection{Performance Analysis by Deception Levels}
\label{sec:modelvshumanresult}

We reassessed each detector's vulnerability to deception using the sample dataset mirroring the user study. We employed each detector's EER equal error point as the discrimination threshold (\autoref{sec:aucandeer}). For detectors with near-random AUC scores, we provide data performance without further discussion. Additionally, we evaluated \texttt{Qwen2-Audio}~\cite{chu2024qwen2}, a multimodal LLM (MLLM), to explore its potential in deepfake voice detection.

We analyzed detectors' FARs on voices graded by user study results, as shown in \autoref{tab:humanmodel}. A higher FAR indicates poorer performance. We found that for both Chinese and English datasets, on level 0 deepfake voices which are difficult to deceive humans, most detectors' FARs are higher than human performance. This means that deepfake voices easily identified as fake by humans are more likely to deceive the detectors, potentially misleading humans in practical applications. For the English dataset, we noticed that except for \texttt{RawBoost} and \texttt{RawGAT-ST}, detectors' FARs increased with deepfake voice levels, though not significantly. 
For level 1 and level 2 deepfake voices, the detectors' performance was notably better than humans'.
For the Chinese dataset, all detectors showed a significant FAR increase from level 0 to 1, but less pronounced from level 1 to 2. Notably, \texttt{AASIST2}, \texttt{RawBMamba}, \texttt{RawGAT-ST}, and \texttt{RawNet2-Vocoder} performed better on level 2 than level 1. 
At levels 1 and 2, all detectors outperformed humans, but at level 1, most detectors' FARs (40\%-50\%) were not significantly better than human performance, except \texttt{AASITS2} (27.17\% FAR).
In summary, most detectors performed poorly on deepfake voices that are difficult to deceive humans. While they provide some assistance in detecting deepfake voices that humans find difficult to judge, the best FAR achieved was just 25.64\%.

The MLLM, \texttt{Qwen-audio2}, exhibits relatively consistent FARs across various deception levels in both languages (\autoref{tab:humanmodel}). However, its performance is poor, with an F1-score of only 33.16\% on the Chinese dataset and 0\% on the English dataset. 
Interestingly, while MLLMs demonstrate superior capabilities in voice analysis, they struggle with detection tasks. This gap reveals that analysis proficiency doesn't guarantee deepfake voice detection capability.

\subsection{Focus Analysis}% 
\label{sec:focus analysis}

\noindent \textbf{Factors Affecting Human Judgment.} 
Based on the participants' ratings of the quality of deepfake voices in the questionnaire (see \autoref{sec:studydesign}), we analyze the influencing factors chosen by humans when judging deepfake voices of varying deception levels.
\autoref{fig:wordcloud} in the Appendix shows the word clouds created using TF-IDF weights for the influencing factors at different levels.
We found that the influencing factors provided by participants did not vary significantly across different scores. High-frequency factors like speech rate, emotion, pauses, and breathing were consistently highlighted in both English and Chinese datasets, while background noise, volume, interjections, and laughter received less attention. Timbre was also frequently mentioned as a key attribute. 
The timbre attribute is usually associated with specific speakers. In the voice collection process (see \autoref{sec:datasetconstruction}), it cannot be guaranteed that all timbres are unfamiliar to the participants. Participants may have encountered some of these timbres on social media or through other channels, and this familiarity could potentially help them in judging deepfake voices.

\noindent \textbf{Divergence in Feature Recognition: Models vs. Humans.}
Detectors typically use specific feature extractors (such as \texttt{OC-Softmax}) or data-driven deep learning models (such as \texttt{AASIST2}, which uses wav2vec 2.0) to learn features that distinguish deepfake voices from real voices. However, the features learned by these models do not always effectively reflect attributes that humans focus on, such as emotions.
Deepfake voices that are easily recognizable by humans often exhibit noticeable differences from real voices in these human-perceived attributes. However, unless the detector's feature extraction can capture these attributes (e.g., the wav2vec 2.0 is widely applied in emotion recognition tasks), the detector may struggle to distinguish such deepfake voices from real voices, as these deepfake voices might resemble the average performance of voice features the model focuses on.
Conversely, for high-quality deepfake voices that are difficult for humans to identify, although they may be similar to real voices in human-perceived attributes, the model's feature extractor might capture subtler differences than human perception.
This may explain the interesting phenomenon in \autoref{sec:modelvshumanresult}: most detectors perform poorly when handling low-quality deepfake voices that are easily identifiable by humans but tend to outperform humans when dealing with high-quality deepfake voices that are challenging to discern.

\section{Discussion}

\subsection{Threats to Validity}%

\noindent \textbf{Overlap Algorithms.} During voice collection (see \autoref{sec:voices collection}), we are unaware of the underlying algorithms used by different commercial tools. As a result, their algorithms may overlap with the open-source models we identified. Without public disclosure, avoiding this overlap is challenging. Fortunately, our experimental results do not indicate significant issues.

\noindent \textbf{Different Sample Rates.} Our dataset (including subsets under resample manipulation) has a different sample rate (see \autoref{sec:datastd}) from the training dataset, which could lead to significant performance differences for sampling rate-sensitive detectors compared to the results reported in the original paper, thereby affecting our evaluation. 

\noindent \textbf{Different Voice Duration.} Differences in voice duration between our evaluation and training datasets (see \autoref{sec:evamethods}) could impact detector performance. However, in real-world scenarios, data duration is unpredictable and should not be an excuse for performance decline.

\noindent \textbf{Participants' Random Choices.} Despite adding attention tests to our user study (as mentioned in \autoref{sec:studydesign}), it is still impossible to completely prevent participants from making random choices for certain deepfake voices, which could affect the grading of deepfake voices. However, our statistical analysis of the F1-Scores and ACCs (see \autoref{sec:userstudyresult}) across three experimental rounds showed that the results were very consistent, indicating that our user study effectively reflected the participants' judgment ability.

\subsection{Diverse Data Challenges}
Datasets with a wide range of generation methods are crucial. Existing deepfake voice datasets, such as ASVspoof2021, excel in this by including voices generated by hundreds of algorithms. However, they share a common issue: a bias towards a single language and a lack of consideration for real-world voice manipulations. Our evaluation shows that detectors trained on a single language suffer significant performance drops with languages outside their training set and degrade further when facing certain manipulations. This suggests that detectors performing well on specific datasets may struggle with real-world deepfake voice detection. Although our work considers a variety of manipulations, it cannot cover all potential voice variations in practical scenarios. In an era where people with diverse native languages can easily communicate and where generating deepfake voices is low-cost, it is crucial to consider the diversity of deepfake voice datasets. Developing detection methods that do not rely on language-specific features and enhancing detector robustness against various manipulations are vital for real-world deployment.

\subsection{Human Focus: Key to Deepfake Detection}

We categorized deepfake voices into three levels based on their ability to deceive humans, aiming to differentiate their quality. Results on the English dataset show that detector performance declines as the deception level increases, though the overall decline is not significant. Interestingly, detectors are more easily fooled by deepfake voices with low deception levels for humans, while they perform better on those with high deception levels. This suggests that when humans can accurately judge deepfake voices, relying solely on detectors may yield misleading results. Human judgment of deepfake voices is a complex process. Although we did not explore this process in detail, we asked participants to identify factors influencing their decisions. \textbf{The findings reveal common focus areas—such as emotion, speech rate, and pauses—when humans evaluate deepfake voices.} The ultimate goal of detectors is to help humans avoid deception. Future research should explore how to better align detector performance with human judgment, whether incorporating common human judgment features can reduce detectors' errors for low-deception deepfakes, and how to enhance overall detector effectiveness.

\section{Conclusion}
\label{sec:conclusion}

Our benchmark \bench{} fills the current gap in systematic and intuitive evaluation for deepfake voice detection. We collected a large set of deepfake voices in English and Chinese using a wide range of commercial tools and advanced open-source methods. Through manipulation, we addressed the limitations of existing datasets that are restricted to a single language or lack variation.
We demonstrated the performance differences of various detectors under different manipulations, revealing that even the most advanced detectors still face significant challenges in real-world applications.
We also conducted a large-scale user study and found that detectors might mislead humans when dealing with deepfake voices of low deception, and identified the key features humans rely on to distinguish deepfake voices.

\bibliographystyle{plain}
\bibliography{main}

\clearpage
\appendix
\section*{APPENDIX}
\label{sec:appendix}
\subsection{Ethics and Artifact Availability }
\label{sec:ethics}
\noindent\textbf{Ethics.} As mentioned in \autoref{sec:voices collection}, for each commercial tool, we carefully examined their terms of service to ensure they can be used for research purposes. We paid the necessary usage fees and ensured that \bench{} is non-commercial, thereby protecting their intellectual property. Before recruiting participants, we obtained the approval from our institution as an exempt study. We only collected participants' gender and age, without any identifiable or sensitive information, qualifying for exemption. As stated in \autoref{sec:studydesign}, our instructions clearly informed participants of the requirements and time limits, and they were free to withdraw at any time without restrictions. 

\noindent \textbf{Artifact Availability.} 
Our artifact including dataset, user study results, weighted models for experimental evaluation, and original outputs can be accessed through our leaderboard (available at \url{https://voicewukong.github.io/}).

\subsection{Participants} 
\label{sec:Participants}

All participants were at least 18 years old and had a minimum of an undergraduate degree. We ensured that each participant was fluent in either English or Chinese to prevent language unfamiliarity from affecting their judgment of deepfake voices. Each participant received \$5 for their participation. A total of 318 participants successfully completed the questionnaire within the specified time limit: 114 completed the Chinese questionnaire and 204 completed the English version. The participant pool had an average age of 22.40 years, comprising 64.47\% males and 35.53\% females. Detailed information about the participants for both Chinese and English questionnaires is shown in \autoref{tab:Participants}.

\begin{figure}[h]
    \centering
    \subfloat[Task 1: Whether the voice \\is human-generated?]{
        \includegraphics[width=0.22\textwidth]{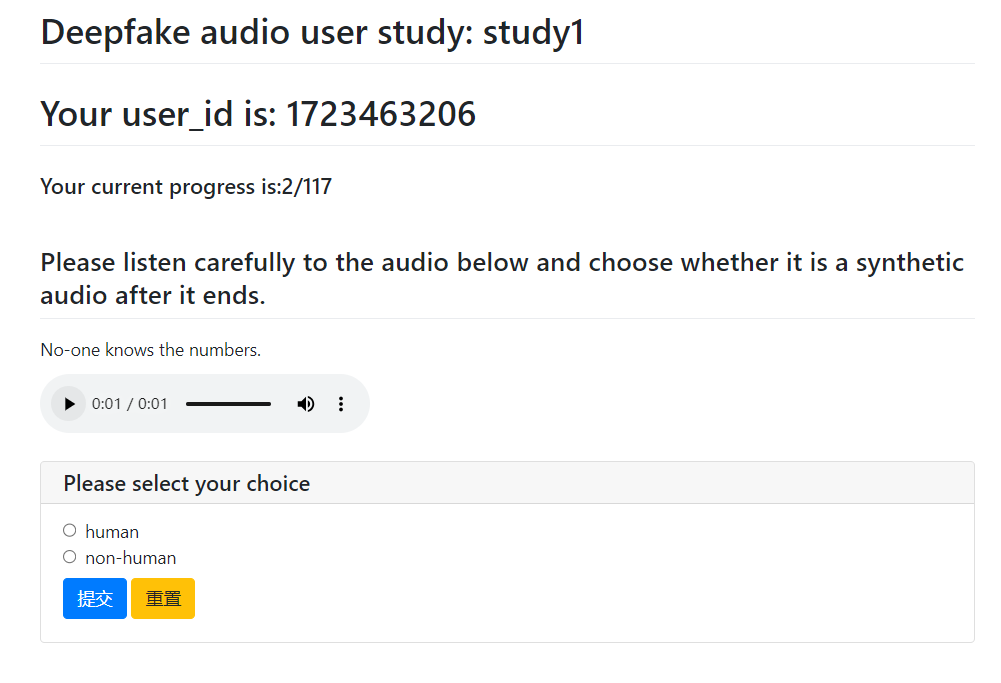}
        \label{fig:task1}
    }
    \subfloat[Task2-1: Rate the generation quality of the deepfake voice.]{
        \includegraphics[width=0.22\textwidth]{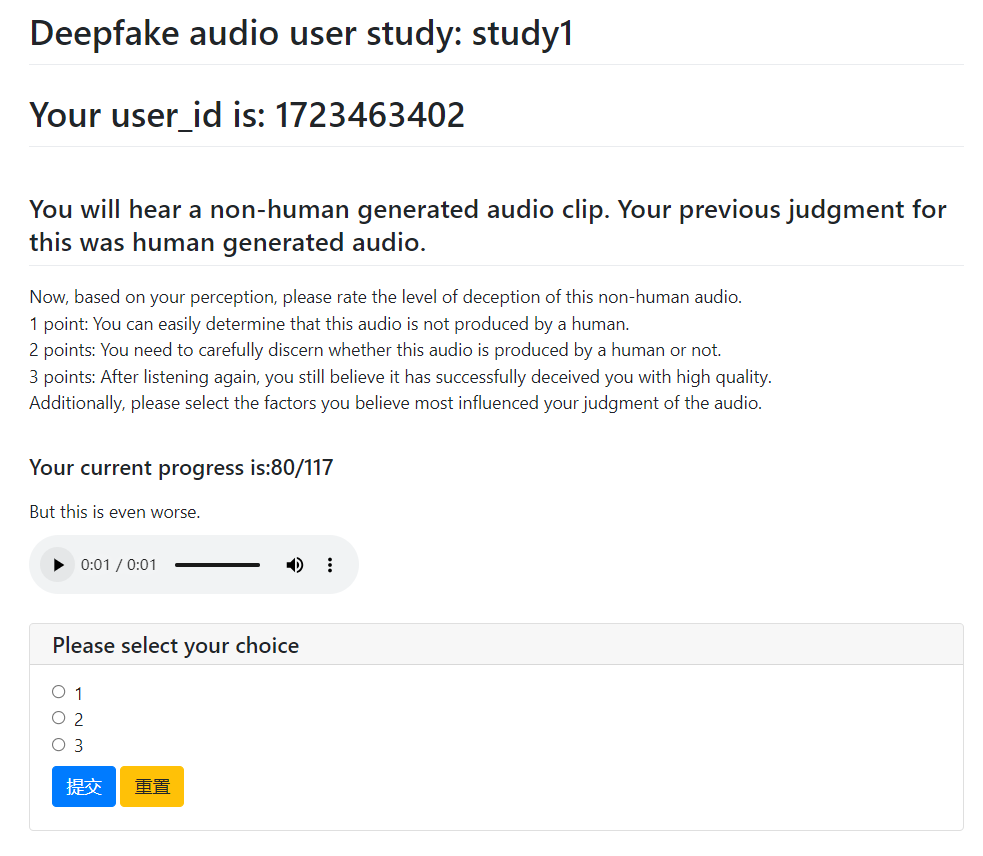}
        \label{fig:task3}
    }
    
    \subfloat[Task2-2:What are the factors that influenced your judgment?]{
        \includegraphics[width=0.22\textwidth]{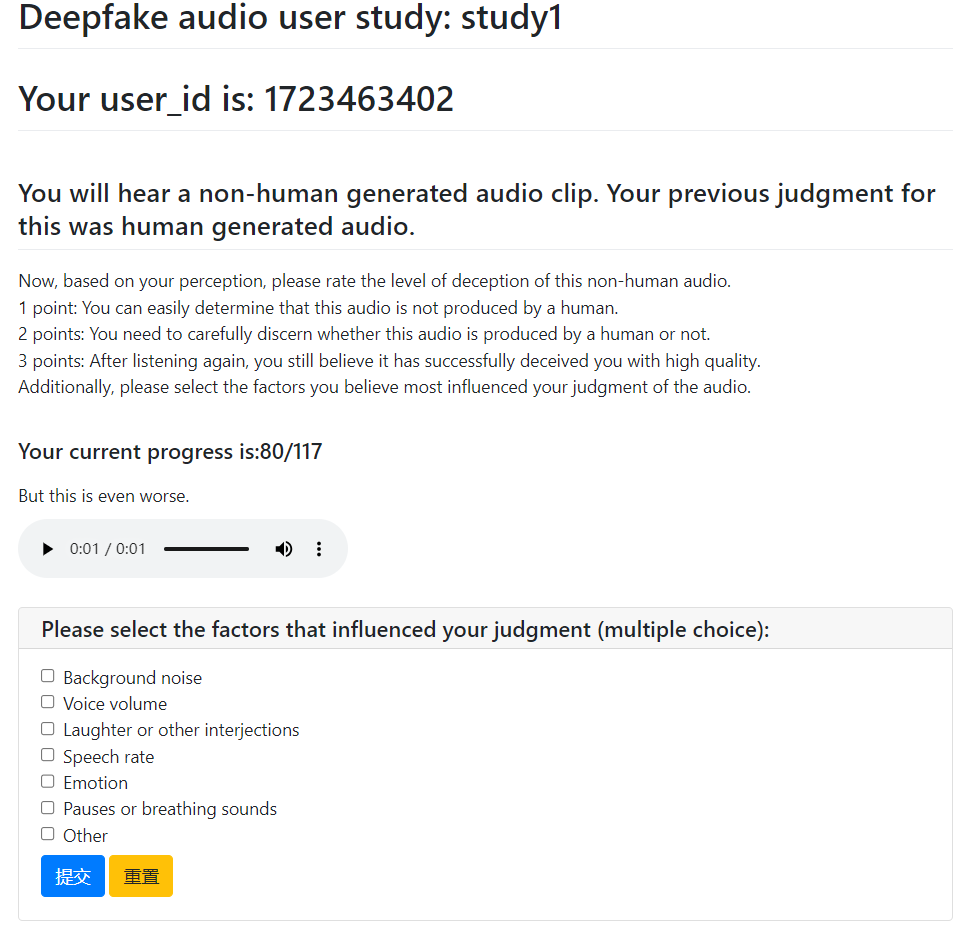}
        \label{fig:task3b}
    }
    \subfloat[Task 3: Attention test.]{
        \includegraphics[width=0.22\textwidth]{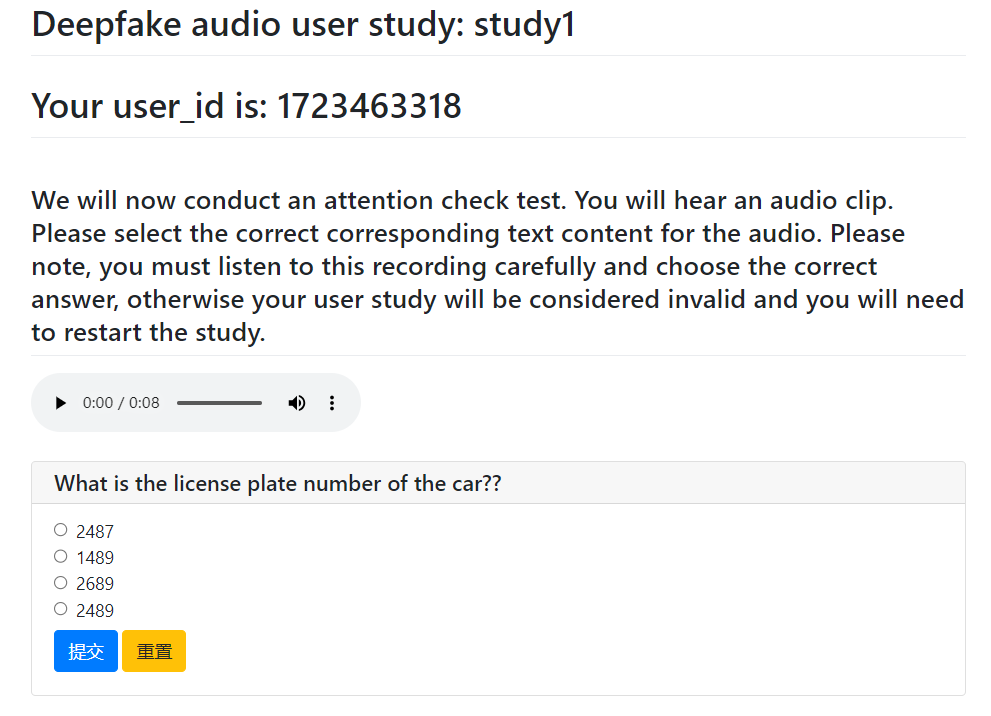}
        \label{fig:task2}
    }
    \caption{Examples of the three tasks in the questionnaire. The figure shows examples from the English questionnaire. The tasks in the Chinese questionnaire are the same, only the language changed.}
    \label{fig:questionnaire}
\end{figure}

\begin{table}[h]
\centering
\caption{Details of participants grouped by English and Chinese language. The average age of the English group was 22.15, while for the Chinese group it was 22.84. The maximum age for both groups was 29 years old. The minimum age for the English group was 18 years old, and for the Chinese group, it was 20 years old.}
\label{tab:Participants}
\resizebox{0.85\linewidth}{!}{%
\begin{tabular}{|r|l|r|l|}
\hline
\multicolumn{2}{|c|}{\textbf{English}} & \multicolumn{2}{c|}{\textbf{Chinese}} \\
\hline
 \bf Sample size & $204$ &  \bf Sample size & $114$ \\
\hline
 \bf $\text{Age}_{\textit{avg}}$(SD) & $22.15$ ($1.89$) &  \bf $\text{Age}_{\textit{avg}}$(SD) & $22.84$ ($2.93$) \\
\hline
 \bf $\text{Age}_{\textit{max}}$ & $29$ &  \bf $\text{Age}_{\textit{max}}$ & $29$ \\
\hline
 \bf $\text{Age}_{\textit{min}}$ & $18$ &  \bf $\text{Age}_{\textit{min}}$ & $20$ \\
\hline
 \bf Female (\%) & $34.8\%$ &  \bf Female (\%) & $36.84\%$ \\
\hline
 \bf Male (\%) & $65.20\%$ &  \bf Male (\%) & $63.16\%$ \\
\hline
 \bf Non-binary (\%)  & 0 &  \bf Non-binary (\%)  & 0\\
\hline
\end{tabular}%
}
\end{table}

\begin{figure}[ht!]
    \centering
    \includegraphics[width=0.95\linewidth]{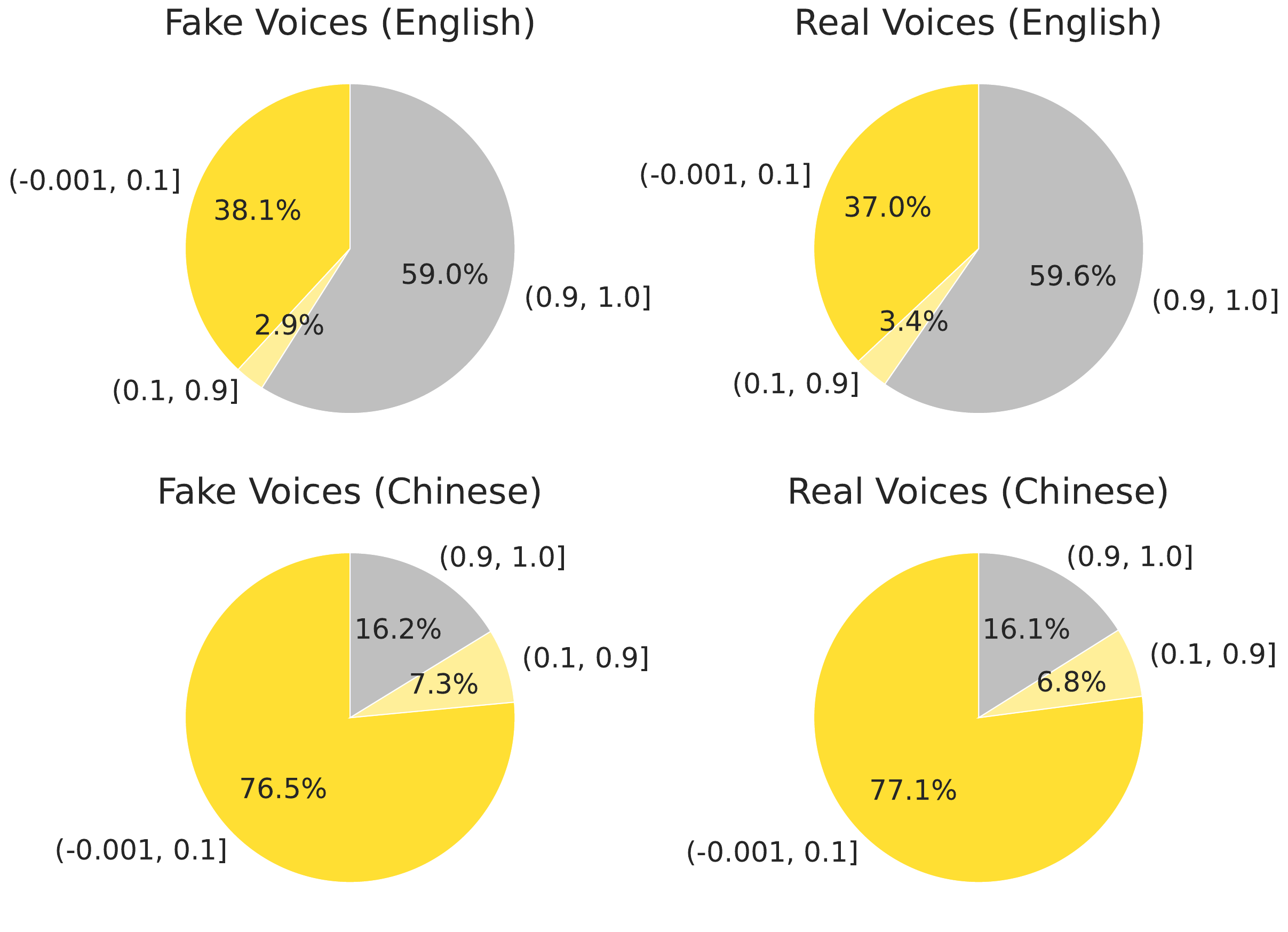}
    \caption{The distribution of evaluation scores for \texttt{Res-TSSDNet} on the std. subset.}
    \label{fig:tssdeval}
\end{figure}

\subsection{Futher discussion about \texttt{Res-TSSDNet}} 
\label{sec:restssdnet}

\texttt{Res-TSSDNet}  has an ACC of less than 50\%, which is the lowest among all the detectors. Its ACC is also just over 50\% (50.86\%) on the Chinese dataset, as shown in \autoref{fig:tssdeval}. The prediction scores of \texttt{Res-TSSDNet} for deepfake voices and real voices are concentrated in two separate, disconnected score ranges on both language datasets, which indicates it lacks generalization capability and the ability to distinguish deepfake voices on our dataset, resulting in an ACC close to random prediction.

\begin{table*}[ht!] 
\caption{The list of generation methods. It includes 19 commercial tools and 15 open-source tools, among which 19 methods support Chinese deepfake voice generation. 29 are dedicated to TTS synthesis and 5 specialize in VC.}
\label{tab:genmethod}
\centering
\footnotesize
\setlength{\tabcolsep}{4pt}
\begin{tabularx}{\textwidth}{@{\extracolsep{\fill}}*{4}{>{\centering\arraybackslash}X} p{8cm}@{}}
\toprule
\textbf{Method} & \textbf{Languages} & \textbf{Fake Types} &\textbf{Commercial} & \textbf{Link} \\

\midrule
XTTS\_v2 & English \& Chinese & TTS & & \url{https://github.com/coqui-ai/TTS} \\
OpenVoice~\cite{qin2023openvoice} & English \& Chinese & TTS & & \url{https://github.com/myshell-ai/OpenVoice} \\
Open-AI API & English & TTS & \checkmark & \url{https://platform.openai.com/docs/guides/text-to-speech} \\
Tortoise-TTS~\cite{betker2023tortoise} & English &  TTS & & \url{https://github.com/neonbjb/tortoise-tts} \\
StyleTTS\_v2~\cite{li2023styletts2humanleveltexttospeech} & English & TTS & & \url{https://github.com/yl4579/StyleTTS2?tab=readme-ov-file} \\

VALL-E X~\cite{zhang2023vallex} & English &  TTS & & \url{https://github.com/Plachtaa/VALL-E-X} \\
Unmuxr & English \& Chinese & TTS & \checkmark & \url{https://app.unmixr.com/studio} \\
Elevenlabs & English & TTS & \checkmark & \url{https://elevenlabs.io/app/speech-synthesis} \\

Play & English & TTS & \checkmark & \url{https://play.ht/} \\

GPT-Sovits & English \& Chinese & TTS & & \url{https://github.com/RVC-Boss/GPT-SoVITS} \\

Vits~\cite{kim2021vitstts} & English & TTS & & \url{https://github.com/jaywalnut310/vits/tree/main} \\
VoiceMaker & English \& Chinese & TTS & \checkmark & \url{https://voicemaker.in/} \\
Voiser & English \& Chinese & TTS & \checkmark & \url{https://voiser.net/account/text-to-speech/} \\
Textmagic & English & TTS & \checkmark & \url{https://freetools.textmagic.com/text-to-speech} \\
Bark & English \& Chinese & TTS & & \url{https://github.com/suno-ai/bark?tab=readme-ov-file} \\
FreeTTS & English \& Chinese & TTS & \checkmark & \url{https://freetts.com/} \\
Voxify & English \& Chinese & TTS & \checkmark & \url{https://voxify.ai/} \\
AiVOOV & English \& Chinese & TTS & \checkmark & \url{https://aivoov.com/} \\
speakperfect  & English & TTS & \checkmark & \url{https://speakperfect.co/} \\
Diffgan-TTS~\cite{liu2022diffganttshighfidelityefficienttexttospeech} & English  & TTS & & \url{https://github.com/keonlee9420/DiffGAN-TTS} \\
Fakeyou & English  & TTS & \checkmark & \url{https://fakeyou.com/tts} \\
ChatTTS & English \& Chinese & TTS & & \url{https://github.com/2noise/ChatTTS} \\
DDDM-VC~\cite{choi2024dddm} & English  & VC &  & \url{https://github.com/hayeong0/DDDM-VC.git} \\
Diff-HierVC~\cite{choi2023diffhiervcdiffusionbasedhierarchicalvoice} & English  & VC & & \url{https://github.com/hayeong0/Diff-HierVC} \\
QuickVC~\cite{guo2023quickvc} & English  & VC & & \url{https://github.com/quickvc/QuickVC-VoiceConversion} \\
StarGan\_v2~\cite{li2021starganv2vcdiverseunsupervisednonparallel} & English \& Chinese & VC & & \url{https://github.com/Ydoit/StarGANv2-VC} \\
RVC~\cite{li2021starganv2vcdiverseunsupervisednonparallel} & English \& Chinese & VC & & \url{https://huggingface.co/lj1995/VoiceConversionWebUI} \\
Resemble-AI & English  & TTS & \checkmark & \url{https://app.resemble.ai/} \\
TextToVoice & English \& Chinese  & TTS & \checkmark & \url{https://www.texttovoice.online/} \\
AnyToSpeech & English  & TTS & \checkmark & \url{https://anytospeech.com/} \\

Audyo & English \& Chinese & TTS & \checkmark & \url{https://www.audyo.ai/} \\
NaturalReader & English \& Chinese & TTS & \checkmark & \url{https://www.naturalreaders.com/online/} \\
Speechgen & English \& Chinese & TTS & \checkmark & \url{https://speechgen.io/} \\
Google API & English \& Chinese & TTS & \checkmark & \url{https://cloud.google.com/text-to-speech} \\
\bottomrule
\end{tabularx}
\end{table*}
\vfill

\begin{figure*}[!t]
    \centering
    \begin{subfigure}{0.9\linewidth}
        \centering
        \includegraphics[width=\linewidth]{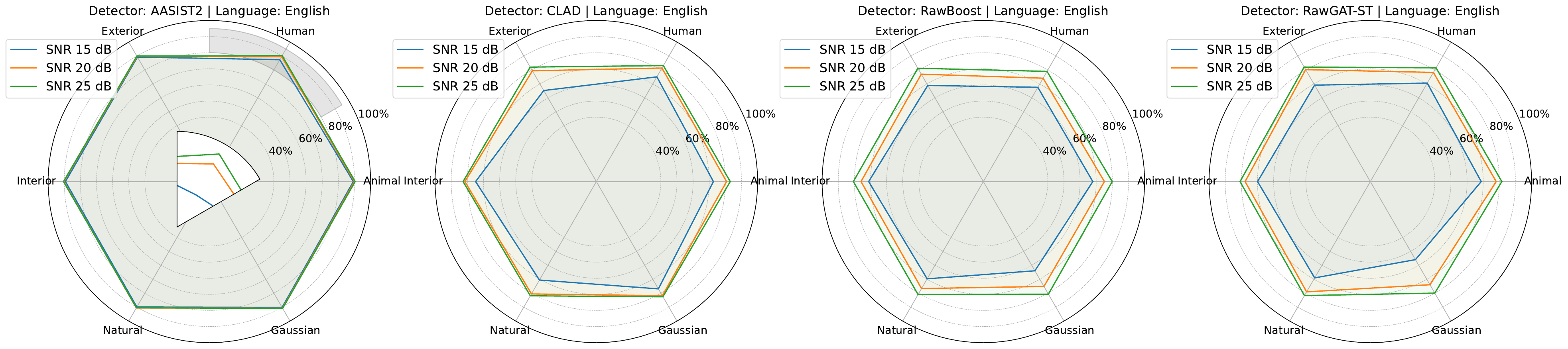}
       
        \caption{ACCs of the top four AUC-ranked detectors under noise injection at different SNR levels on the English dataset.}
         \label{fig:ennoise}
    \end{subfigure}
    \hfill
    \begin{subfigure}{0.9\linewidth}
        \centering
        \includegraphics[width=\linewidth]{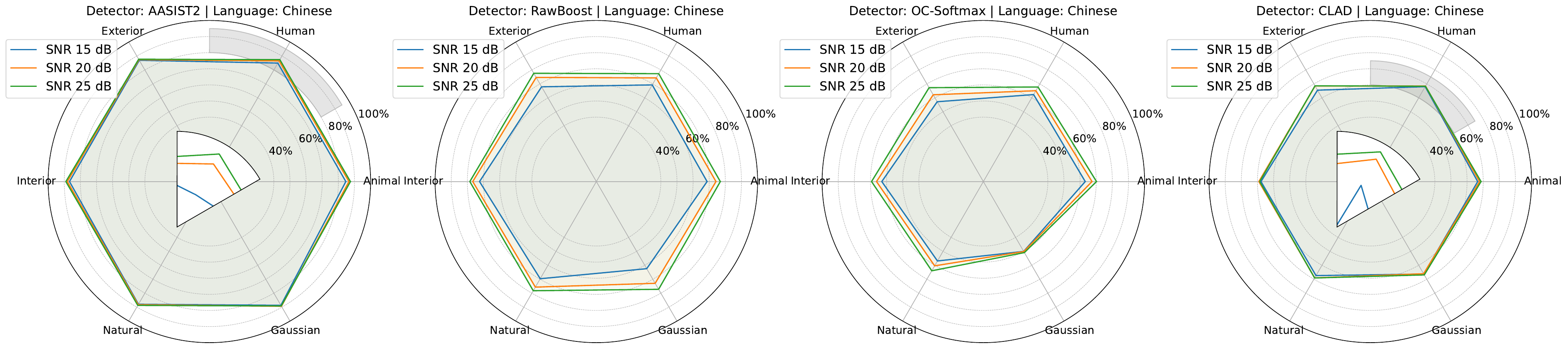}
        
        \caption{ACCs of the top four AUC-ranked detectors under noise injection at different SNR levels on the Chinese dataset.}
        \label{fig:zhnoise}
    \end{subfigure}
    \caption{ACCs of the top four AUC-ranked detectors under noise injection at different SNR levels.}
    \label{fig:noisesnr}
\end{figure*}

\begin{figure}[htbp]
    \centering
    \includegraphics[width=\linewidth]{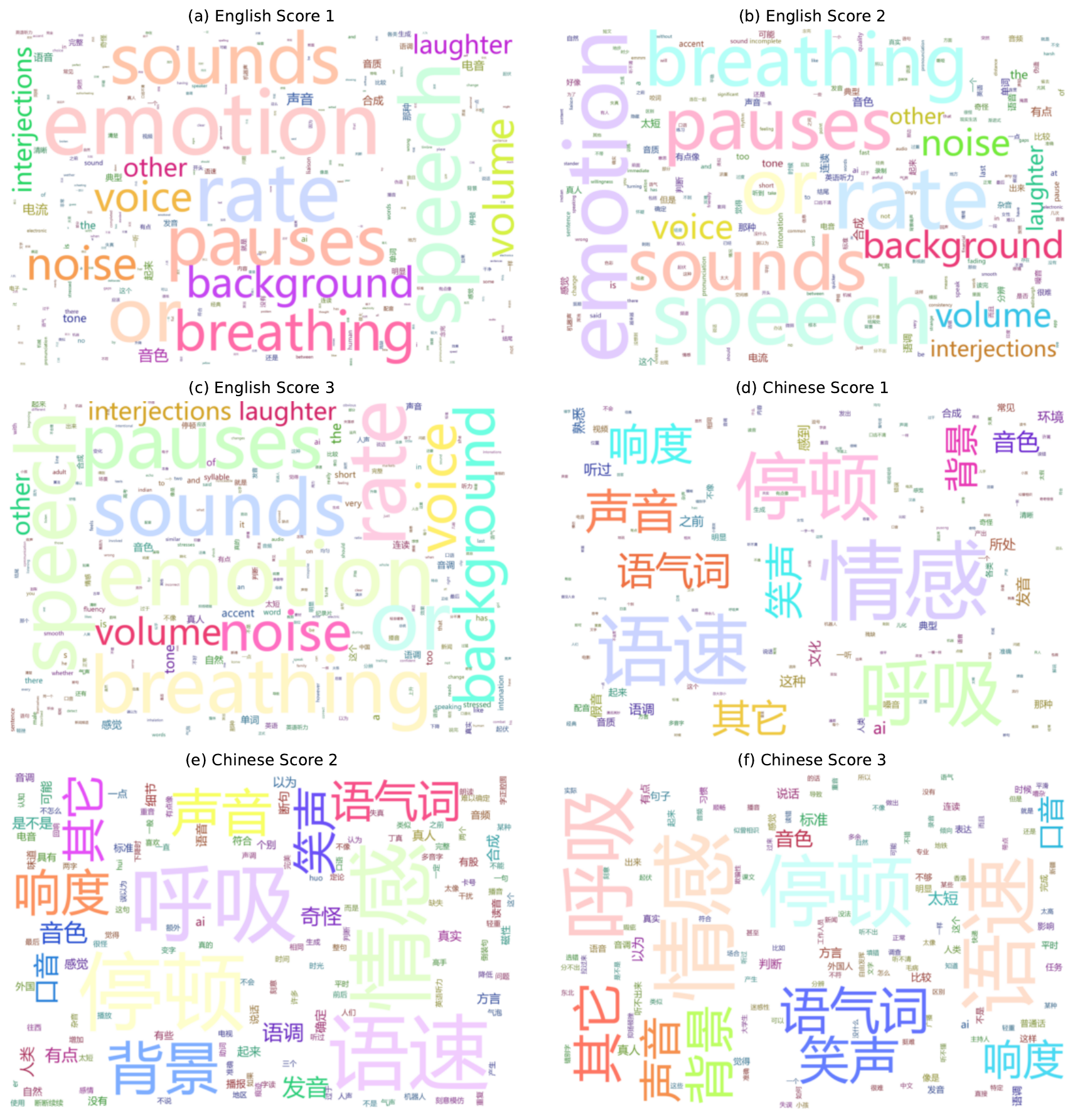}
    \caption{Word clouds of influencing factors given by participants, weighted by TF-IDF. Subplots (a), (b), and (c) show results from the English dataset, while the remaining figures present results from the Chinese dataset.}
    \label{fig:wordcloud}
\end{figure}

\begin{figure}[htbp]
    \centering
    \includegraphics[width=\linewidth]{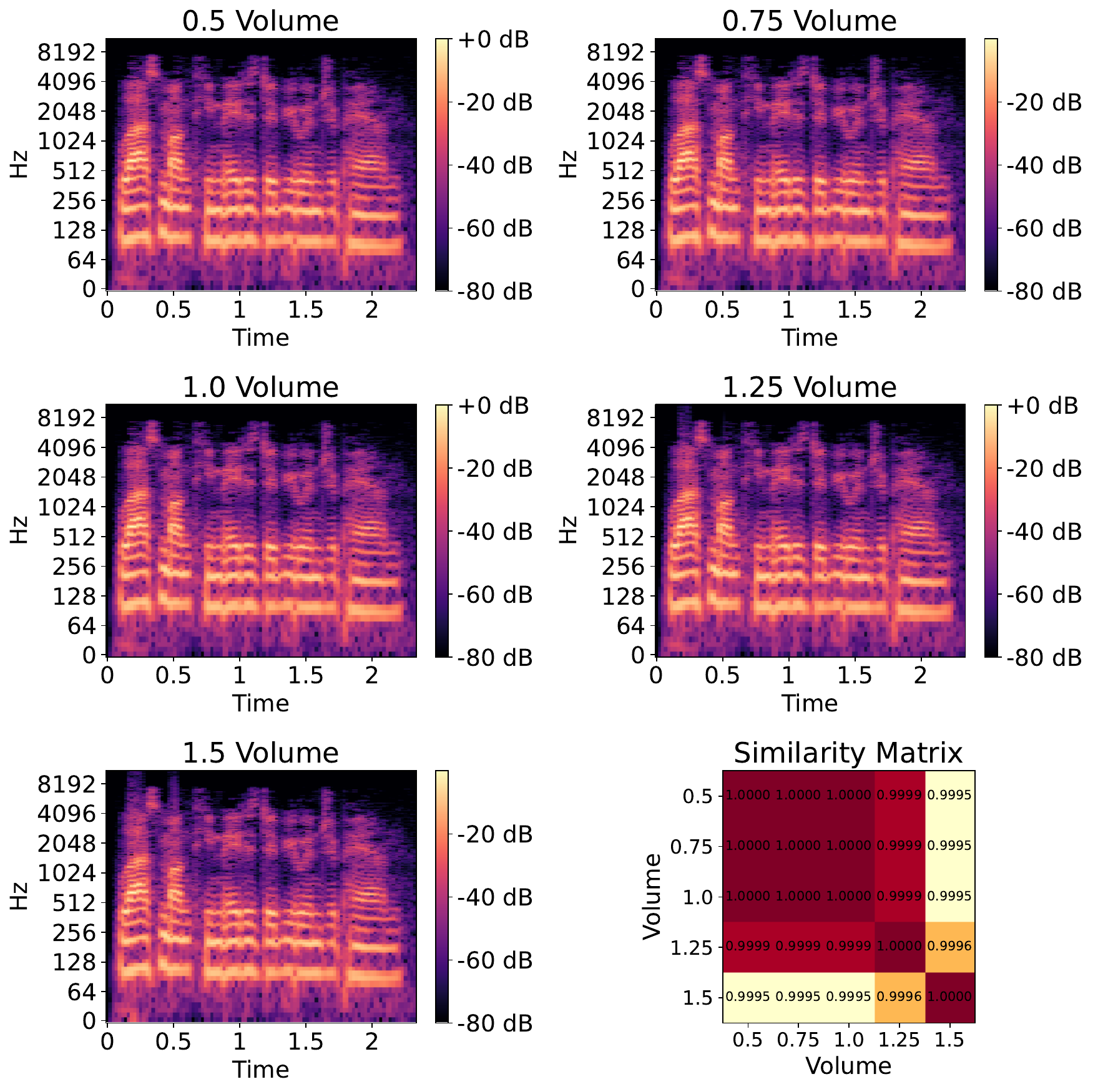}
    \caption{Spectrograms of the same voice at different volume levels and the cosine similarity between these spectrograms. The spectrograms' similarity of voices with different volume levels remains extremely high.}
    \label{fig:spectrogram_similarity}
\end{figure}

\begin{table*}[ht!] 
\caption{Detectors' ACCs (\%) and F1-Scores (\%) on various subsets in the English dataset. In the Noise Injection (NI) manipulation, ``G'' represents \textit{Gaussian}, ``N'' represents \textit{Natural}, ``I'' represents \textit{Interior}, ``E'' represents \textit{Exterior}, ``H'' represents \textit{Human}, and ``A'' represents \textit{Animal}. In the Fade In \& Out (FD) manipulation, ``LE'' represents linear fade, ``E'' represents exponential fade, and ``LG'' represents logarithmic fade.}

\label{tab:accandf1en}
\centering

\tiny
\setlength{\tabcolsep}{4pt}
\begin{tabularx}{\textwidth}{X|X|XXXX|XXXX|XXXX} % 

\hline
\vfill   \textbf{Types}   & \vfill \centering \textbf{Metrics}  & \vfill \centering  \textbf{AASIST2} &  \vfill \centering \textbf{RowBoost} & \vfill \centering \textbf{CLAD} & \vfill \centering \textbf{RawGAT-ST} & \vfill  \centering \textbf{OC-Softmax} & \vfill \centering \textbf{RawNet2} & \vfill \centering\textbf{RawBmamba} &  \centering \textbf{Raw PC-DARTS} & \centering \textbf{RawNet2-Vocoder} & \vfill  \centering \textbf{SAMO} & \vfill \centering \textbf{AASIST} &  \multicolumn{1}{c}{\raisebox{-1.9ex}{\textbf{Res-TSSDNet}}}  \\
\hline
\multirow{2}{*}{\makecell[c]{Std.}} & \makecell[c]{ACC}  & \makecell[c]{91.24}  & \makecell[c]{83.69}  & \makecell[c]{85.25}  & \makecell[c]{82.35}  & \makecell[c]{80.43}  & \makecell[c]{79.9}  & \makecell[c]{83.05}  & \makecell[c]{84.78}  & \makecell[c]{81.4}  & \makecell[c]{81.93}  & \makecell[c]{78.51}  & \makecell[c]{49.48}  \\
 & \makecell[c]{F1-Score}  & \makecell[c]{91.67}  & \makecell[c]{83.83}  & \makecell[c]{86.37}  & \makecell[c]{83.48}  & \makecell[c]{82.52}  & \makecell[c]{80.08}  & \makecell[c]{83.07}  & \makecell[c]{86.72}  & \makecell[c]{81.15}  & \makecell[c]{83.79}  & \makecell[c]{81.17}  & \makecell[c]{14.51}  \\ 
\hline
\multirow{2}{*}{\makecell[c]{NI(G-15dB)}} & \makecell[c]{ACC}  & \makecell[c]{90.19}  & \makecell[c]{63.95}  & \makecell[c]{76.92}  & \makecell[c]{55.96}  & \makecell[c]{50.0}  & \makecell[c]{72.07}  & \makecell[c]{53.45}  & \makecell[c]{51.63}  & \makecell[c]{54.58}  & \makecell[c]{60.79}  & \makecell[c]{55.91}  & \makecell[c]{49.72}  \\
 & \makecell[c]{F1-Score}  & \makecell[c]{90.81}  & \makecell[c]{68.03}  & \makecell[c]{73.23}  & \makecell[c]{28.21}  & \makecell[c]{66.67}  & \makecell[c]{76.95}  & \makecell[c]{68.24}  & \makecell[c]{67.4}  & \makecell[c]{67.13}  & \makecell[c]{38.89}  & \makecell[c]{22.75}  & \makecell[c]{15.22}  \\
\hline

\end{tabularx}
\end{table*}

\begin{table*}[ht!]
\caption{Detectors' ACCs (\%) and F1-Scores (\%) on various subsets in the Chinese dataset. The subsets in this table are consistent with those in \autoref{tab:accandf1en}.}
\label{tab:accandf1zh}
\centering

\tiny
\setlength{\tabcolsep}{4pt}
\begin{tabularx}{\textwidth}{X|X|XXXX|XXXX|XXXX}

\hline
% \hline
\vfill   \textbf{Types}   & \vfill \centering \textbf{Metrics}  & \vfill \centering  \textbf{AASIST2} &  \vfill \centering \textbf{RowBoost} & \vfill \centering \textbf{CLAD} & \vfill \centering \textbf{RawGAT-ST} & \vfill  \centering \textbf{OC-Softmax} & \vfill \centering \textbf{RawNet2} & \vfill \centering\textbf{RawBmamba} &  \centering \textbf{Raw PC-DARTS} & \centering \textbf{RawNet2-Vocoder} & \vfill  \centering \textbf{SAMO} & \vfill \centering \textbf{AASIST} &  \multicolumn{1}{c}{\raisebox{-1.9ex}{\textbf{Res-TSSDNet}}}  \\
\hline
\multirow{2}{*}{\makecell[c]{Std.}} & \makecell[c]{ACC}  & \makecell[c]{90.07}  & \makecell[c]{79.54}  & \makecell[c]{68.68}  & \makecell[c]{71.45}  & \makecell[c]{81.16}  & \makecell[c]{70.67}  & \makecell[c]{76.22}  & \makecell[c]{81.38}  & \makecell[c]{70.53}  & \makecell[c]{51.47}  & \makecell[c]{48.72}  & \makecell[c]{50.86}  \\
 & \makecell[c]{F1-Score}  & \makecell[c]{90.24}  & \makecell[c]{78.5}  & \makecell[c]{70.5}  & \makecell[c]{72.06}  & \makecell[c]{82.56}  & \makecell[c]{68.89}  & \makecell[c]{75.47}  & \makecell[c]{83.5}  & \makecell[c]{67.6}  & \makecell[c]{55.16}  & \makecell[c]{52.73}  & \makecell[c]{50.9}  \\ 
\hline
\multirow{2}{*}{\makecell[c]{NI(G-15dB)}} & \makecell[c]{ACC}  & \makecell[c]{88.7}  & \makecell[c]{62.45}  & \makecell[c]{66.29}  & \makecell[c]{61.04}  & \makecell[c]{50.04}  & \makecell[c]{60.87}  & \makecell[c]{50.28}  & \makecell[c]{51.53}  & \makecell[c]{45.55}  & \makecell[c]{52.49}  & \makecell[c]{49.59}  & \makecell[c]{49.86}  \\
 & \makecell[c]{F1-Score}  & \makecell[c]{89.15}  & \makecell[c]{70.12}  & \makecell[c]{60.22}  & \makecell[c]{50.06}  & \makecell[c]{66.68}  & \makecell[c]{68.95}  & \makecell[c]{66.55}  & \makecell[c]{67.24}  & \makecell[c]{61.14}  & \makecell[c]{35.37}  & \makecell[c]{25.37}  & \makecell[c]{48.89}  \\
\hline
\end{tabularx}
\end{table*}

\end{document}